\documentclass[reprint,superscriptaddress,amssymb,amsmath,aps,showpacs,10pt,floatfix,prb,longbibliography]{revtex4-2}
\def\Tb{TbMn$_6$Sn$_6$}
\def\Gd{GdMn$_6$Sn$_6$}
\def\Re{$R$Mn$_6$Sn$_6$}

\def\FeSn{Fe$_3$Sn$_2$}
\def\cm{cm$^{-1}$}

\usepackage{graphicx}%
\usepackage{color}
\usepackage{epstopdf}
\usepackage{amssymb}
\usepackage{amsmath}
\usepackage{amsfonts}
\usepackage[note-name=, use-sort-key = false]{notes2bib}

\usepackage{xcolor}

\usepackage{color}
\usepackage[colorlinks,bookmarks=false,citecolor=darkblue,linkcolor=red,urlcolor=blue]{hyperref} 
\definecolor{darkred}{rgb}{0.7,0.0,0.0}

\definecolor{darkblue}{rgb}{0,0.02,0.45}

\definecolor{darkgreen}{rgb}{0.02,0.45,0.0}

\definecolor{violet}{rgb}{0.8,0.2,0.6}

\begin{document}
\title{Effect of magnetism and phonons on localized carriers in the ferrimagnetic kagome metals \Gd\ and \Tb}

\author{M. Wenzel}
\email{maxim.wenzel@pi1.physik.uni-stuttgart.de}
\affiliation{1. Physikalisches Institut, Universit{\"a}t Stuttgart, 70569 
Stuttgart, Germany}

\author{A. A. Tsirlin}
\affiliation{Felix Bloch Institute for Solid-State Physics, Leipzig University, 04103 Leipzig, Germany}
\affiliation{Experimental Physics VI, Center for Electronic Correlations and Magnetism, Institute of Physics, University of Augsburg, 86135 Augsburg, Germany}

\author{O. Iakutkina}
\affiliation{1. Physikalisches Institut, Universit{\"a}t Stuttgart, 70569 
Stuttgart, Germany}

\author{Q. Yin}
\affiliation{Laboratory for Neutron Scattering, and Beijing Key Laboratory of Optoelectronic Functional Materials MicroNano Devices, Department of Physics, Renmin University of China, Beijing 100872, China}

\author{H.C. Lei}
\affiliation{Laboratory for Neutron Scattering, and Beijing Key Laboratory of Optoelectronic Functional Materials MicroNano Devices, Department of Physics, Renmin University of China, Beijing 100872, China}

\author{M. Dressel}
\affiliation{1. Physikalisches Institut, Universit{\"a}t Stuttgart, 70569 
Stuttgart, Germany}

\author{E. Uykur}
\affiliation{1. Physikalisches Institut, Universit{\"a}t Stuttgart, 70569 
Stuttgart, Germany}
\affiliation{Helmholtz-Zentrum Dresden-Rossendorf, Institute of Ion Beam Physics and Materials Research, 01328 Dresden, Germany}

\date{\today}

\begin{abstract}
Kagome metals possess peculiar optical spectra consisting of contributions from free charge carriers in a Drude-type response, localized carriers seen as a strongly temperature-dependent localization peak, and, in some cases, phonons displaying strong anomalies. The rare-earth kagome metal series, \Re, provides a marvelous playground to study the electronic properties of kagome metals in the presence of variable magnetic order. Here, we report temperature-dependent reflectivity studies on two members of the \Re\ family, \Gd\ (in-plane ferrimagnet) and \Tb\ (out-of-plane ferrimagnet), in a broad energy range (50~-~18000~\cm, equivalent to 6.2~meV~-~2.23~eV) down to 10~K. At high temperatures, a phonon mode at approximately 160~\cm\ is observed, which becomes screened out in \Tb\ below $\sim$~150~K as the localization peak linearly passes through the mode. In \Gd, the disappearance of the phonon is accompanied by the onset of saturation of the peak position, suggesting an unusual interplay between the two features.
\end{abstract}

\maketitle

Proposed by Sy\^ozi in 1951, the kagome lattice quickly became popular among both theoretical and experimental physicists due to its unique magnetic and electronic properties \cite{Syozi1951, Mekata2003}. Consisting of spatially separated metallic kagome planes, kagome metals are model compounds for studying strong electronic correlations, magnetism, and topologically non-trivial states \cite{Liu2019}. Here, the itinerant carriers give rise to the peculiar kagome electronic band structure hosting dispersionless flat bands, saddle points, as well as linearly dispersing Dirac bands \cite{Wang2013, Lin2018, Kang2020, Li2021, Liu2020, Liu2021}.

\begin{figure*}
    \hspace{-4mm}
    \begin{minipage}{0.25\linewidth}
        \centering
       \includegraphics[width=1\columnwidth]{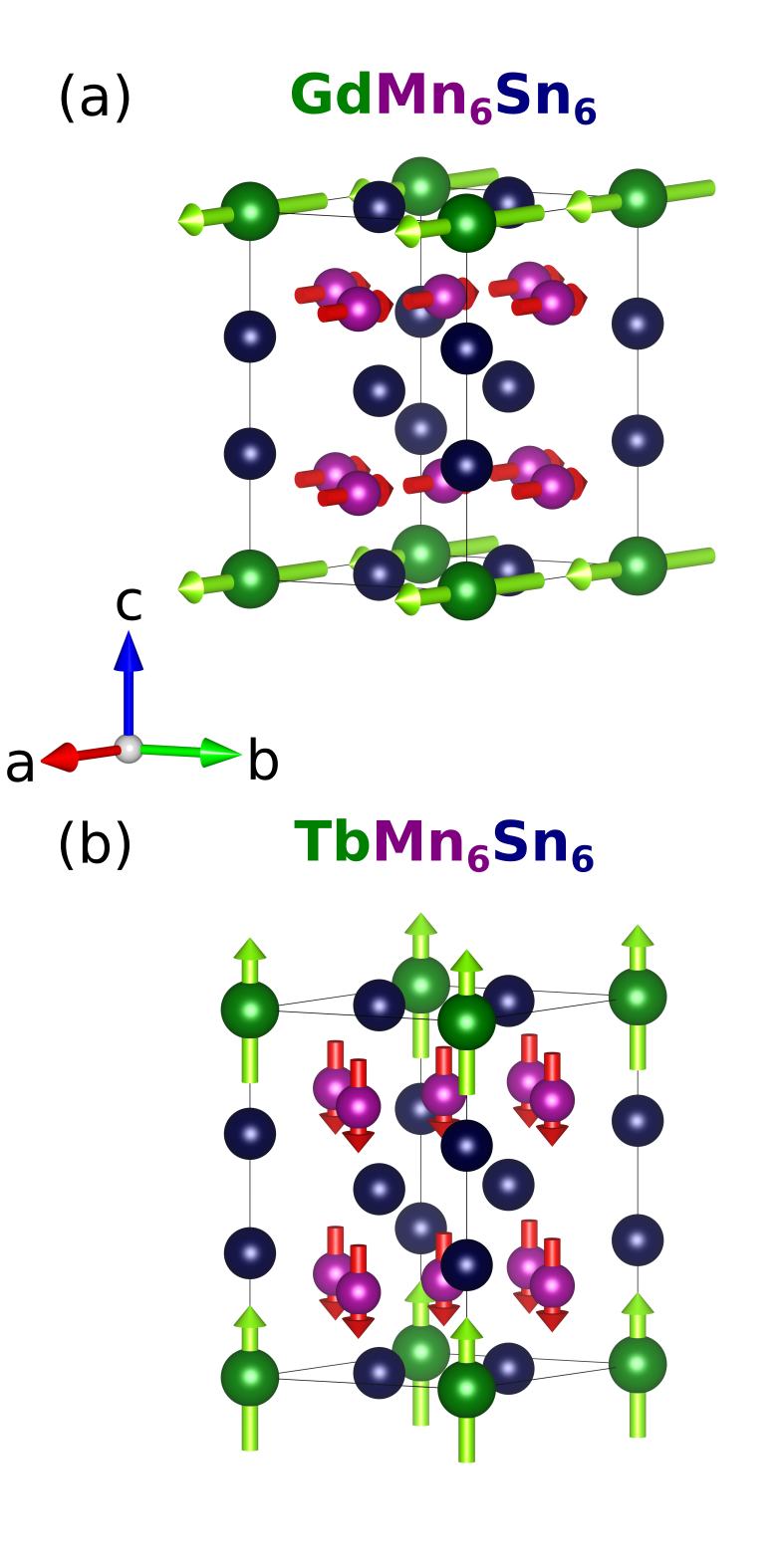}
    \end{minipage}
    \begin{minipage}{0.75\linewidth}
        \centering
        \includegraphics[width=1\columnwidth]{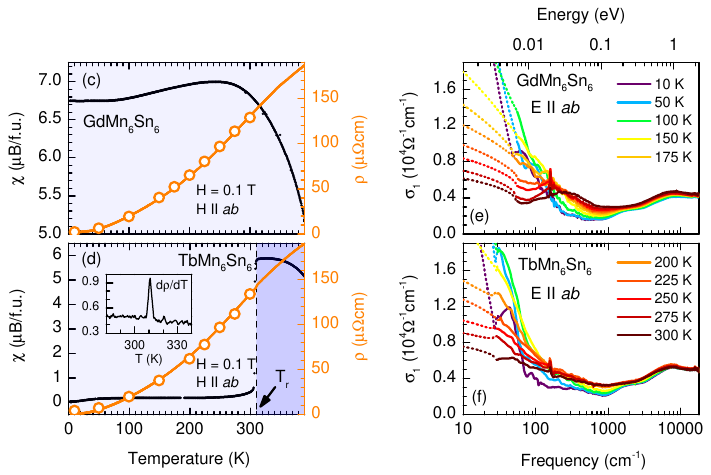}
    \end{minipage}
   \caption{(a) and (b) Crystal and magnetic structure below 300~K of \Gd\ and \Tb, respectively \cite{Malaman1999, Idrissi1991}. (c) and (d) Magnetic susceptibility and dc resistivity curves measured in the $ab$-plane. The Curie temperature of both systems lies above the measured temperature range; however, a spin reorientation from the basal plane near to the $c$-axis around $T_{\mathrm{r}}\sim$~310~K is visible for \Tb. For \Gd, no anomalies are observed in the measured temperature range. Open circles are the dc resistivity values obtained from the Hagen-Rubens fits of the optical measurements as explained in the Supplemental Material~\cite{SM}. (e) and (f) Temperature-dependent in-plane optical conductivity with the dotted lines being the Hagen-Rubens extrapolation to low energies.}	
    \label{spectrum}
\end{figure*}

The ternary rare-earth series, \Re\ ($R$ = Sc, Y, Gd-Lu), opens new ways to investigate the influence of magnetism on the electronic properties of kagome metals and hence, distinguish between magnetic-driven and kagome layer-driven properties. While these compounds have been studied extensively over the last three decades regarding their unusual magnetic structure, they recently gained attention in the framework of kagome metals \cite{Venturni1991, Ghimire2020, Yin2020}. These compounds crystallize in the $P6/mmm$ space group featuring spatially decoupled magnetic Mn-kagome planes stacked along the $c$-axis, which are stabilized by Sn1 atoms. Within one unit cell, the kagome layers are separated by non-magnetic Sn2 atoms forming a honeycomb lattice, while $R$Sn3 layers separate the kagome planes from one unit cell to another, as sketched in Figs.~\ref{spectrum}(a) and \ref{spectrum}(b). The underlying magnetic structure strongly depends on the rare-earth element separating the layers, resulting in a large variety of ferrimagnetic ($R$ = Gd, Tb, Dy, Ho) and antiferromagnetic ($R$ = Sc, Y, Er, Tm, Yb, Lu) ground states across the series \citep{Venturni1991, Malaman1999}.

Angle-resolved photoemission spectroscopy (ARPES) and Landau level measurements reveal the signatures of the kagome lattice, including topologically non-trivial Dirac bands and flat bands in these materials \cite{Gu2022, Li2021, Liu2021, Yin2020}. Comprising spin-polarized Mn 3$d$ states with a strong intrinsic spin-orbit coupling, these two-dimensional kagome bands exhibit non-trivial Chern numbers \cite{Kang2020, Li2021, Xu2015} giving rise to an intrinsic anomalous Hall effect \cite{Ma2021, Xu2021, Xu2022, Gao2021, Asaba2020, Dhakal2021}. While the different magnetic structures do not seem to affect the main band dispersions near the Fermi energy $E_{\mathrm{F}}$, significantly, a gap at the Dirac points has been proposed only for the ferrimagnetic systems \cite{Yin2020, Ma2021a, Haldane1988, Sims2022}. Moreover, this Chern gap, as well as the energy of the Dirac points $E_{\mathrm{D}}$, can be tuned with the rare-earth element \cite{Ma2021a}. Here, the number of unpaired 4$f$ electrons of the rare-earth element plays an important role as a coupling between the 4$f$ and the 3$d$ electrons is observed. 

The key implications of these topological features lie in unusual transport properties that crucially rely on charge carriers and their dynamics \cite{Yin2020, Ortiz2019, Neupert2022, Zhang2022}. Especially the effect of magnetism is one of the central issues \cite{Cheng2022}. Therefore, here, we study these dynamics and their dependence on the magnetic order with temperature-dependent broadband Fourier transform infrared spectroscopy studies on \Re\ systems, namely on \Gd\ and \Tb. While both systems possess an almost identical crystal structure and a ferrimagnetic ground state below room temperature, in the former one, the spins are aligned within the kagome plane, whereas in the Tb compound, a perpendicular alignment to the kagome layers is reported \cite{Venturni1991, Malaman1999, Yazdi2011, Jones2022, Mielke2022}. This was confirmed prior to our optical study by dc transport and magnetic susceptibility measurements shown in Figs.~\ref{spectrum}(c) and \ref{spectrum}(d). We further performed density functional theory plus Hubbard $U$ (DFT+$U$) calculations to evaluate the electronic structures, revealing the correlated character of the \Re\ series. Due to localization effects, the optical response of the charge carriers splits into the conventional Drude part and a prominent low-energy peak. This peak shows a clear dependence on the magnetic order and underlies the magnetic tunability of this compound family. 

\begin{figure*}
\centering
	\includegraphics[width=2\columnwidth]{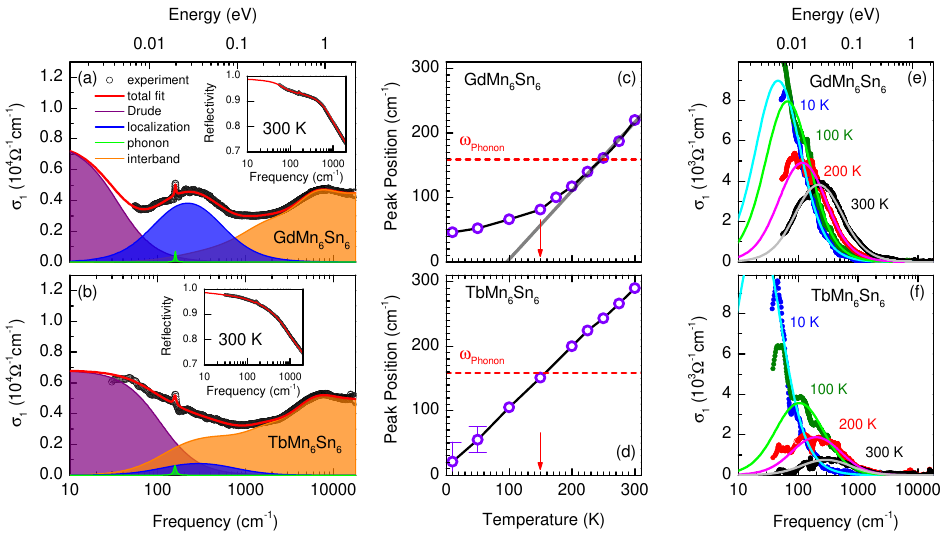}
	\caption{(a) and (b) Decomposed optical conductivity at 300~K, consisting of a Drude component (purple), a localization peak (blue), a phonon mode (green), and several interband transitions (orange). The insets show the total fit to the measured reflectivity. Details on the fitting process as well as the decomposed spectra at lower temperatures can be found in the Supplemental Material~\cite{SM}. (c) and (d) Temperature dependence of the localization peak position. The red dashed line marks the phonon mode, while the red arrow indicates the temperature where the mode disappears. (e) and (f) Temperature evolution of the localization peak, obtained by subtracting the fitted Drude, phonon mode, and interband contributions from the spectra. The solid lines are the Fratini model fits to the total experimental conductivity as described in the Supplemental Material~\cite{SM}.}		
	\label{interplay}
\end{figure*}

Figures~\ref{spectrum}(e) and \ref{spectrum}(f) display the temperature-dependent real part of the in-plane optical conductivity of \Gd\ and \Tb, respectively. At first glance, the spectra are remarkably similar and resemble the spectrum of the ferromagnetic \FeSn\ \cite{Biswas2020, Schilberth2021}. Consistent with the metallic nature of these compounds, a very narrow Drude component is observed at low energies, which becomes even sharper upon cooling. For \Gd, only the tail of this feature is visible even at 300~K. Two step-like absorption features can be identified in the otherwise relatively flat conductivity at high energies. Very similar steps were interpreted as the signature of two-dimensional Dirac fermions in \FeSn. In addition to the sharp Drude component and interband transitions, a phonon mode around 160 \cm\ is observed. 
Furthermore, we have realized that the low-energy dynamics cannot be reproduced only with a single Drude component, but an additional peak-like absorption feature is required as shown in Fig.~\ref{interplay} (a) and (b) for the data at 300~K. With this peak showing a strong red-shift upon cooling, it puts the \Re\ series on common ground with other kagome metals and clearly separates this feature from other low-energy transitions, which are interband in nature \cite{Biswas2020, Wenzel2022, Uykur2022, Uykur2021}.

A closer look at the low-energy regime reveals substantial differences between the two ferrimagnetic compounds. Figures~\ref{interplay} (b) and \ref{interplay}(d) show the temperature evolution of this so-called localization peak in \Gd\ and \Tb\ after subtracting the fitted Drude, phonon, and interband contributions from the experimental optical conductivity. Not only is the localization peak more pronounced in the in-plane ferrimagnetic system \Gd, but the peak position saturates at low temperatures, as shown in Fig.~\ref{interplay}(a). In contrast, a linear red-shift over the whole temperature range is observed in \Tb\ [see Fig.~\ref{interplay}(c)]. Hence, the peak moves out of the measured range at low temperatures, and its position has to be estimated from its high-frequency tail, as well as by considering the linear behavior of the shift at higher temperatures, leading to increasing error bars of the fits. 

Visually, the temperature evolution of the peak position in \Gd\ looks strikingly
similar to the behavior in \FeSn. For the latter, a possible coupling between the localization peak and the underlying magnetic structure is discussed since
the linear scaling breaks down after a reorientation of the Fe-spins at 120 K \cite{Biswas2020, Kumar2019}. Additionally, the shape of the peak transforms into a sharp Fano resonance. The saturation as observed in \Gd\ was also reported in the non-magnetic KV$_3$Sb$_5$, suggesting that the origin of this effect may be other than magnetic. Additionally, no change of the in-plane ferrimagnetic structure of \Gd\ is reported below room temperature; hence, the primary cause for the observed saturation must be something else. Nevertheless, a commonality between the two magnetic systems is the in-plane direction of the magnetic moments in both \FeSn\ below its spin-reorientation transition and \Gd.

One plausible explanation for the observed saturation uniting magnetic and non-magnetic kagome metals is the involvement of a phonon mode. Indeed, phonons and their importance for the electronic structure of kagome metals have been studied in multiple compounds. In the $A$V$_3$Sb$_5$ family, phonons are discussed to be the driving force behind the charge-density-wave formation and the low-temperature superconductivity \cite{Luo2022, Zhong2022}. Optical measurements revealed strong phonon anomalies associated with a coupling of the phonon modes to the electronic background in KV$_3$Sb$_5$ and RbV$_3$Sb$_5$ \cite{Wenzel2022, Uykur2022}. Furthermore, a strong interplay between phonons and fermionic degrees of freedom was revealed by scanning tunneling microscopy (STM) studies of paramagnetic CoSn \cite{Yin2020a}.

DFT calculations, shown in the Supplemental Material~\cite{SM}, reveal a total number of nine IR-active phonon modes in each compound. Four of these modes have the A$_{2u}$ symmetry involving out-of-plane atomic displacements and hence, cannot be detected by our in-plane measurements. While in highly metallic systems phonon modes are often too weak to be detected and/or screened by the free carriers, our measurements were able to capture a prominent E$_{1u}$ mode around 160~\cm\ at room temperature. At low temperatures, this mode disappears in both compounds. At first glance, this anomalous behavior might be explained by a structural distortion; however, low-temperature XRD studies report almost no changes in the crystal structure of \Re\ down to 2~K \cite{Malaman1999, Idrissi1991}. Hence, an interplay between the phonon mode and the localization peak has to be considered as a possible scenario, not least because both features are located around the same energy range. 

For a further comparison of the two features, the position of the phonon mode is marked with the red dashed line in Figs.~\ref{interplay}(a) and \ref{interplay}(c), while the red arrow points at the temperature at which the phonon mode disappears in each compound. In \Tb, the phonon mode disappears as soon as the localization peak passes through it, suggesting that the localization peak screens out the phonon mode. On the other hand, a more complex relationship between the two features is observed in \Gd. Here, the phonon mode shows an enhancement and a slight broadening as the localization peak passes through it, and is retained even below the crossing over a narrow temperature range. Eventually, the mode disappears around the temperature where the position of the localization peak saturates. This behavior suggests an unusual coupling between the phonon mode and the localization peak in \Gd. Based on the observation that the strong localization peak anomalies appear in the in-plane ferromagnetic system, one plausible explanation would be a magneto-elastic coupling to the in-plane infrared-active phonon mode. Additionally, the rare-earth element could directly influence the phonon mode and hence its interplay with the localization peak.

Ultimately, an interplay with some other bosonic excitations such as magnons, for instance, could as well lead to the distinct behavior of the localization peak in \Gd\ compared to \Tb. Indeed, magnon bands extending to energies up to $\sim$~100~meV have been reported in several members of the \Re\ family \cite{Riberolles2022, Zhang2020}. 

\begin{figure}
\centering
	\includegraphics[width=1\columnwidth]{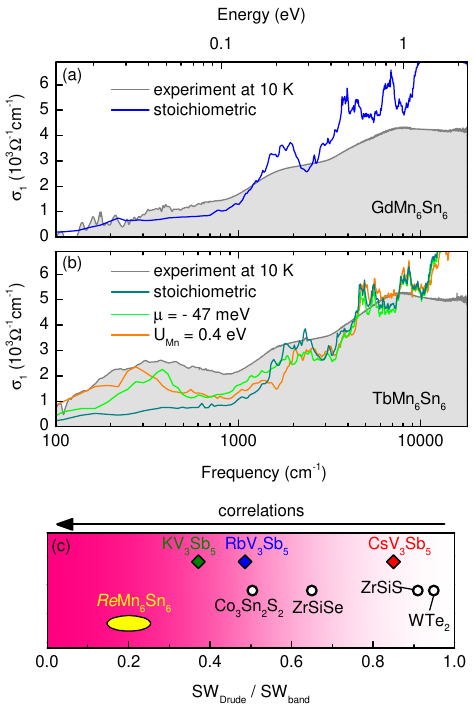}
	\caption{(a) and (b) Experimental interband transitions along with the DFT+$U$ calculated optical conductivity. For all calculations a Hubbard $U_{R} = 10$~eV was added to the rare-earth element. Furthermore, the energy scale of the calculated conductivity is rescaled for a better comparison with the experiment. (c) Correlation scaling for different kagome metals and other topological materials taken from ref.~\cite{Shao2020}.}		
	\label{calculations}
\end{figure}

The presence of a red-shifting localization peak is a common occurrence in systems with slow electron dynamics, such as organic conductors, cuprates, and manganites \cite{Fratini2021, Delacretaz2017}, many of them being strongly correlated materials. Hence, we now turn to analyzing the electronic correlations in the \Re\ series. Figures~\ref{calculations}(a) and \ref{calculations}(b) show the comparison between the experimental and the calculated optical conductivities using DFT taking into account the different magnetic structures. For all calculations, a Hubbard $U_{R}\,=\,10$~eV was added to the rare-earth element with the DFT+$U$ method using the double-counting correction in the fully localized limit to treat the strongly correlated 4$f$ electrons \cite{Petersen2006, Soderlind2014, Lee2022, Liu2021}. In the case of \Gd, a good agreement with the experiment is found, while for \Tb, the low-energy spectral weight cannot be reproduced with this method. The agreement is improved by adding a Hubbard $U_{\mathrm{Mn}}$~=~0.4~eV to the Mn-atoms. Another possibility is shifting the Fermi energy down by 47~meV; however, this requires removing one electron from the structure, which is hard to reconcile with the system. 

Although with different adjustments, one can bring the calculations to the experiment's level, in either case, the energy of the calculated conductivity needs to be rescaled by a factor of 2.5 in \Gd\ (2 in \Tb). A very similar scaling factor was previously reported for ARPES measurements of \Gd\ \cite{Liu2021}. This suggests that these systems are clearly beyond DFT, and electronic correlations therein can not be fully treated on the mean-field DFT+$U$ level.

We further observed the step-like absorption features, combined with the relatively flat optical conductivity, as the potential signatures of the Dirac points in these systems. Considering that there are two Dirac points, one above and one below the Fermi energy (see Supplemental Material~\cite{SM}), one would expect these step-like absorption features to appear \cite{Biswas2020}. This interpretation becomes even more tempting when the energies of the steps are compared with the ARPES measurements. However, considering the relatively high energy range of these features and the significant number of bands crossing the Fermi energy, the step-like absorption is most likely just a cumulative effect of different contributions; hence, one should be careful in its assignment. On the other hand, absorption features at lower energies ($\omega < 1000$ \cm) can be related to transitions between bands very close to the Fermi energy, most probably involving transitions between the saddle points nearby the $M$ point, as shown in our band structure calculations in the Supplemental Material~\cite{SM}.

Although the \Re\ series lies beyond the limits of the DFT+$U$ methods presented here, the calculations can be used for an initial assessment of the correlation strength. As proposed previously for different compounds, including cuprates, iron pnictides, and topologically nontrivial Dirac systems \cite{Shao2020, Quazilbash2009}, the ratio of the spectral weight of the mobile carriers from the experiment and the DFT calculations can be used as a gauge of electronic correlations. Here, SW$_{\mathrm{Drude}}$/SW$_{\mathrm{band}}$ is close to 1 for uncorrelated materials, while the ratio becomes zero for Mott insulators showing the most correlated behavior. Figure~\ref{calculations}(c) depicts this scaling for the $A$V$_3$Sb$_5$ series and topological semimetals taken from refs.~\cite{Shao2020, Wenzel2022}. From the calculations, we can determine a rough value of SW$_{\mathrm{Drude}}$/SW$_{\mathrm{band}}\,\approx\,0.2$, pointing towards much stronger correlations in comparison with the $A$V$_3$Sb$_5$ series and other kagome metals reported to date. Moreover, no significant difference between \Gd\ and \Tb\ is observed, whereas the correlation strength changes drastically between different members of the $A$V$_3$Sb$_5$ family.

In summary, we establish the correlated nature of ferrimagnetic kagome metals of the \Re\ family and uncover partial localization of charge carriers manifested by the prominent low-energy peak in the optical conductivity. The temperature evolution of this peak is sensitive to details of the magnetic order. While in \Tb, the localization peak red-shifts linearly through the whole temperature range upon cooling and screens out the phonon mode at $\sim$~160~\cm, it displays different characteristics in \Gd. Here, the peak is more pronounced, while its position saturates at low temperatures. This dissimilar behavior indicates a major difference in low-energy degrees of freedom that damp electron dynamics and, consequently, should affect transport properties at low temperatures. Both compounds display a strongly correlated character, as a good agreement with the experimental interband transitions is only found after rescaling the energy of the calculated optical conductivity, and the experimental Drude spectral weight is drastically lower than the DFT prediction.

The authors acknowledge the fruitful discussion with Simone Fratini, and technical support by Gabriele Untereiner. We also thank Falk Lissner and Rainer Niewa for the XRD measurements. H.C.L. was supported by National Key R\&D Program of China (Grant No. 2018YFE0202600), the Beijing Natural Science Foundation (Grant No. Z200005), the Fundamental Research Funds for the Central Universities and Research Funds of Renmin University of China (RUC) (Grant Nos. 18XNLG14, 19XNLG13, and 19XNLG17), and the Beijing National Laboratory for Condensed Matter Physics. The work has been supported by the Deutsche Forschungsgemeinschaft (DFG) via Grants No. UY63/2-1, No. DR228/48-1, and No. DR228/51-1. E. U. acknowledges the European Social Fund and the Baden-Württemberg Stiftung for the financial support of this research project by the Eliteprogramme.

\nocite{Fratini2014, pbe96, wien2k, Wang2021, Gorbunov2012, Homes1993, Tanner2015, vasp1, vasp2, draxl2006}
\bibliography{TbGd}

\begin{thebibliography}{61}%
\makeatletter
\providecommand \@ifxundefined [1]{%
 \@ifx{#1\undefined}
}%
\providecommand \@ifnum [1]{%
 \ifnum #1\expandafter \@firstoftwo
 \else \expandafter \@secondoftwo
 \fi
}%
\providecommand \@ifx [1]{%
 \ifx #1\expandafter \@firstoftwo
 \else \expandafter \@secondoftwo
 \fi
}%
\providecommand \natexlab [1]{#1}%
\providecommand \enquote  [1]{``#1''}%
\providecommand \bibnamefont  [1]{#1}%
\providecommand \bibfnamefont [1]{#1}%
\providecommand \citenamefont [1]{#1}%
\providecommand \href@noop [0]{\@secondoftwo}%
\providecommand \href [0]{\begingroup \@sanitize@url \@href}%
\providecommand \@href[1]{\@@startlink{#1}\@@href}%
\providecommand \@@href[1]{\endgroup#1\@@endlink}%
\providecommand \@sanitize@url [0]{\catcode `\\12\catcode `\$12\catcode
  `\&12\catcode `\#12\catcode `\^12\catcode `\_12\catcode `\%12\relax}%
\providecommand \@@startlink[1]{}%
\providecommand \@@endlink[0]{}%
\providecommand \url  [0]{\begingroup\@sanitize@url \@url }%
\providecommand \@url [1]{\endgroup\@href {#1}{\urlprefix }}%
\providecommand \urlprefix  [0]{URL }%
\providecommand \Eprint [0]{\href }%
\providecommand \doibase [0]{https://doi.org/}%
\providecommand \selectlanguage [0]{\@gobble}%
\providecommand \bibinfo  [0]{\@secondoftwo}%
\providecommand \bibfield  [0]{\@secondoftwo}%
\providecommand \translation [1]{[#1]}%
\providecommand \BibitemOpen [0]{}%
\providecommand \bibitemStop [0]{}%
\providecommand \bibitemNoStop [0]{.\EOS\space}%
\providecommand \EOS [0]{\spacefactor3000\relax}%
\providecommand \BibitemShut  [1]{\csname bibitem#1\endcsname}%
\let\auto@bib@innerbib\@empty
\bibitem [{\citenamefont {Syôzi}(1951)}]{Syozi1951}%
  \BibitemOpen
  \bibfield  {author} {\bibinfo {author} {\bibfnamefont {I.}~\bibnamefont
  {Syôzi}},\ }\bibfield  {title} {\bibinfo {title} {{Statistics of Kagomé
  Lattice}},\ }\href {https://doi.org/10.1143/ptp/6.3.306} {\bibfield
  {journal} {\bibinfo  {journal} {Progress of Theoretical Physics}\ }\textbf
  {\bibinfo {volume} {6}},\ \bibinfo {pages} {306} (\bibinfo {year}
  {1951})}\BibitemShut {NoStop}%
\bibitem [{\citenamefont {Mekata}(2003)}]{Mekata2003}%
  \BibitemOpen
  \bibfield  {author} {\bibinfo {author} {\bibfnamefont {M.}~\bibnamefont
  {Mekata}},\ }\bibfield  {title} {\bibinfo {title} {{Kagome: The Story of the
  Basketweave Lattice}},\ }\href {https://doi.org/10.1063/1.1564329} {\bibfield
   {journal} {\bibinfo  {journal} {Physics Today}\ }\textbf {\bibinfo {volume}
  {56}},\ \bibinfo {pages} {12} (\bibinfo {year} {2003})}\BibitemShut {NoStop}%
\bibitem [{\citenamefont {Liu}\ \emph {et~al.}(2019)\citenamefont {Liu},
  \citenamefont {Liang}, \citenamefont {Liu}, \citenamefont {Xu}, \citenamefont
  {Li}, \citenamefont {Chen}, \citenamefont {Pei}, \citenamefont {Shi},
  \citenamefont {Mo}, \citenamefont {Dudin}, \citenamefont {Kim}, \citenamefont
  {Cacho}, \citenamefont {Li}, \citenamefont {Sun}, \citenamefont {Yang},
  \citenamefont {Liu}, \citenamefont {Parkin}, \citenamefont {Felser},\ and\
  \citenamefont {Chen}}]{Liu2019}%
  \BibitemOpen
  \bibfield  {author} {\bibinfo {author} {\bibfnamefont {D.~F.}\ \bibnamefont
  {Liu}}, \bibinfo {author} {\bibfnamefont {A.~J.}\ \bibnamefont {Liang}},
  \bibinfo {author} {\bibfnamefont {E.~K.}\ \bibnamefont {Liu}}, \bibinfo
  {author} {\bibfnamefont {Q.~N.}\ \bibnamefont {Xu}}, \bibinfo {author}
  {\bibfnamefont {Y.~W.}\ \bibnamefont {Li}}, \bibinfo {author} {\bibfnamefont
  {C.}~\bibnamefont {Chen}}, \bibinfo {author} {\bibfnamefont {D.}~\bibnamefont
  {Pei}}, \bibinfo {author} {\bibfnamefont {W.~J.}\ \bibnamefont {Shi}},
  \bibinfo {author} {\bibfnamefont {S.~K.}\ \bibnamefont {Mo}}, \bibinfo
  {author} {\bibfnamefont {P.}~\bibnamefont {Dudin}}, \bibinfo {author}
  {\bibfnamefont {T.}~\bibnamefont {Kim}}, \bibinfo {author} {\bibfnamefont
  {C.}~\bibnamefont {Cacho}}, \bibinfo {author} {\bibfnamefont
  {G.}~\bibnamefont {Li}}, \bibinfo {author} {\bibfnamefont {Y.}~\bibnamefont
  {Sun}}, \bibinfo {author} {\bibfnamefont {L.~X.}\ \bibnamefont {Yang}},
  \bibinfo {author} {\bibfnamefont {Z.~K.}\ \bibnamefont {Liu}}, \bibinfo
  {author} {\bibfnamefont {S.~S.~P.}\ \bibnamefont {Parkin}}, \bibinfo {author}
  {\bibfnamefont {C.}~\bibnamefont {Felser}},\ and\ \bibinfo {author}
  {\bibfnamefont {Y.~L.}\ \bibnamefont {Chen}},\ }\bibfield  {title} {\bibinfo
  {title} {{Magnetic Weyl semimetal phase in a Kagomé crystal}},\ }\href
  {https://doi.org/10.1126/science.aav2873} {\bibfield  {journal} {\bibinfo
  {journal} {Science}\ }\textbf {\bibinfo {volume} {365}},\ \bibinfo {pages}
  {1282} (\bibinfo {year} {2019})}\BibitemShut {NoStop}%
\bibitem [{\citenamefont {Wang}\ \emph {et~al.}(2013)\citenamefont {Wang},
  \citenamefont {Li}, \citenamefont {Xiang},\ and\ \citenamefont
  {Wang}}]{Wang2013}%
  \BibitemOpen
  \bibfield  {author} {\bibinfo {author} {\bibfnamefont {W.-S.}\ \bibnamefont
  {Wang}}, \bibinfo {author} {\bibfnamefont {Z.-Z.}\ \bibnamefont {Li}},
  \bibinfo {author} {\bibfnamefont {Y.-Y.}\ \bibnamefont {Xiang}},\ and\
  \bibinfo {author} {\bibfnamefont {Q.-H.}\ \bibnamefont {Wang}},\ }\bibfield
  {title} {\bibinfo {title} {{Competing electronic orders on kagome lattices at
  van Hove filling}},\ }\href {https://doi.org/10.1103/PhysRevB.87.115135}
  {\bibfield  {journal} {\bibinfo  {journal} {Phys. Rev. B}\ }\textbf {\bibinfo
  {volume} {87}},\ \bibinfo {pages} {115135} (\bibinfo {year}
  {2013})}\BibitemShut {NoStop}%
\bibitem [{\citenamefont {Lin}\ \emph {et~al.}(2018)\citenamefont {Lin},
  \citenamefont {Choi}, \citenamefont {Zhang}, \citenamefont {Qin},
  \citenamefont {Yi}, \citenamefont {Wang}, \citenamefont {Li}, \citenamefont
  {Wang}, \citenamefont {Zhang}, \citenamefont {Sun}, \citenamefont {Wei},
  \citenamefont {Zhang}, \citenamefont {Guo}, \citenamefont {Lu}, \citenamefont
  {Cho}, \citenamefont {Zeng},\ and\ \citenamefont {Zhang}}]{Lin2018}%
  \BibitemOpen
  \bibfield  {author} {\bibinfo {author} {\bibfnamefont {Z.}~\bibnamefont
  {Lin}}, \bibinfo {author} {\bibfnamefont {J.-H.}\ \bibnamefont {Choi}},
  \bibinfo {author} {\bibfnamefont {Q.}~\bibnamefont {Zhang}}, \bibinfo
  {author} {\bibfnamefont {W.}~\bibnamefont {Qin}}, \bibinfo {author}
  {\bibfnamefont {S.}~\bibnamefont {Yi}}, \bibinfo {author} {\bibfnamefont
  {P.}~\bibnamefont {Wang}}, \bibinfo {author} {\bibfnamefont {L.}~\bibnamefont
  {Li}}, \bibinfo {author} {\bibfnamefont {Y.}~\bibnamefont {Wang}}, \bibinfo
  {author} {\bibfnamefont {H.}~\bibnamefont {Zhang}}, \bibinfo {author}
  {\bibfnamefont {Z.}~\bibnamefont {Sun}}, \bibinfo {author} {\bibfnamefont
  {L.}~\bibnamefont {Wei}}, \bibinfo {author} {\bibfnamefont {S.}~\bibnamefont
  {Zhang}}, \bibinfo {author} {\bibfnamefont {T.}~\bibnamefont {Guo}}, \bibinfo
  {author} {\bibfnamefont {Q.}~\bibnamefont {Lu}}, \bibinfo {author}
  {\bibfnamefont {J.-H.}\ \bibnamefont {Cho}}, \bibinfo {author} {\bibfnamefont
  {C.}~\bibnamefont {Zeng}},\ and\ \bibinfo {author} {\bibfnamefont
  {Z.}~\bibnamefont {Zhang}},\ }\bibfield  {title} {\bibinfo {title}
  {{Flatbands and Emergent Ferromagnetic Ordering in
  ${\mathrm{Fe}}_{3}{\mathrm{Sn}}_{2}$ Kagome Lattices}},\ }\href
  {https://doi.org/10.1103/PhysRevLett.121.096401} {\bibfield  {journal}
  {\bibinfo  {journal} {Phys. Rev. Lett.}\ }\textbf {\bibinfo {volume} {121}},\
  \bibinfo {pages} {096401} (\bibinfo {year} {2018})}\BibitemShut {NoStop}%
\bibitem [{\citenamefont {Kang}\ \emph {et~al.}(2020)\citenamefont {Kang},
  \citenamefont {Ye}, \citenamefont {Fang}, \citenamefont {You}, \citenamefont
  {Levitan}, \citenamefont {Han}, \citenamefont {Facio}, \citenamefont
  {Jozwiak}, \citenamefont {Bostwick}, \citenamefont {Rotenberg}, \citenamefont
  {Chan}, \citenamefont {McDonald}, \citenamefont {Graf}, \citenamefont
  {Kaznatcheev}, \citenamefont {Vescovo}, \citenamefont {Bell}, \citenamefont
  {Kaxiras}, \citenamefont {van~den Brink}, \citenamefont {Richter},
  \citenamefont {Prasad~Ghimire}, \citenamefont {Checkelsky},\ and\
  \citenamefont {Comin}}]{Kang2020}%
  \BibitemOpen
  \bibfield  {author} {\bibinfo {author} {\bibfnamefont {M.}~\bibnamefont
  {Kang}}, \bibinfo {author} {\bibfnamefont {L.}~\bibnamefont {Ye}}, \bibinfo
  {author} {\bibfnamefont {S.}~\bibnamefont {Fang}}, \bibinfo {author}
  {\bibfnamefont {J.-S.}\ \bibnamefont {You}}, \bibinfo {author} {\bibfnamefont
  {A.}~\bibnamefont {Levitan}}, \bibinfo {author} {\bibfnamefont
  {M.}~\bibnamefont {Han}}, \bibinfo {author} {\bibfnamefont {J.~I.}\
  \bibnamefont {Facio}}, \bibinfo {author} {\bibfnamefont {C.}~\bibnamefont
  {Jozwiak}}, \bibinfo {author} {\bibfnamefont {A.}~\bibnamefont {Bostwick}},
  \bibinfo {author} {\bibfnamefont {E.}~\bibnamefont {Rotenberg}}, \bibinfo
  {author} {\bibfnamefont {M.~K.}\ \bibnamefont {Chan}}, \bibinfo {author}
  {\bibfnamefont {R.~D.}\ \bibnamefont {McDonald}}, \bibinfo {author}
  {\bibfnamefont {D.}~\bibnamefont {Graf}}, \bibinfo {author} {\bibfnamefont
  {K.}~\bibnamefont {Kaznatcheev}}, \bibinfo {author} {\bibfnamefont
  {E.}~\bibnamefont {Vescovo}}, \bibinfo {author} {\bibfnamefont {D.~C.}\
  \bibnamefont {Bell}}, \bibinfo {author} {\bibfnamefont {E.}~\bibnamefont
  {Kaxiras}}, \bibinfo {author} {\bibfnamefont {J.}~\bibnamefont {van~den
  Brink}}, \bibinfo {author} {\bibfnamefont {M.}~\bibnamefont {Richter}},
  \bibinfo {author} {\bibfnamefont {M.}~\bibnamefont {Prasad~Ghimire}},
  \bibinfo {author} {\bibfnamefont {J.~G.}\ \bibnamefont {Checkelsky}},\ and\
  \bibinfo {author} {\bibfnamefont {R.}~\bibnamefont {Comin}},\ }\bibfield
  {title} {\bibinfo {title} {{Dirac fermions and flat bands in the ideal kagome
  metal FeSn}},\ }\href {https://doi.org/10.1038/s41563-019-0531-0} {\bibfield
  {journal} {\bibinfo  {journal} {Nature Materials}\ }\textbf {\bibinfo
  {volume} {19}},\ \bibinfo {pages} {163} (\bibinfo {year} {2020})}\BibitemShut
  {NoStop}%
\bibitem [{\citenamefont {Li}\ \emph {et~al.}(2021)\citenamefont {Li},
  \citenamefont {Wang}, \citenamefont {Wang}, \citenamefont {Yuan},
  \citenamefont {Song}, \citenamefont {Lou}, \citenamefont {Liu}, \citenamefont
  {Huang}, \citenamefont {Liu}, \citenamefont {Lei}, \citenamefont {Yin},\ and\
  \citenamefont {Wang}}]{Li2021}%
  \BibitemOpen
  \bibfield  {author} {\bibinfo {author} {\bibfnamefont {M.}~\bibnamefont
  {Li}}, \bibinfo {author} {\bibfnamefont {Q.}~\bibnamefont {Wang}}, \bibinfo
  {author} {\bibfnamefont {G.}~\bibnamefont {Wang}}, \bibinfo {author}
  {\bibfnamefont {Z.}~\bibnamefont {Yuan}}, \bibinfo {author} {\bibfnamefont
  {W.}~\bibnamefont {Song}}, \bibinfo {author} {\bibfnamefont {R.}~\bibnamefont
  {Lou}}, \bibinfo {author} {\bibfnamefont {Z.}~\bibnamefont {Liu}}, \bibinfo
  {author} {\bibfnamefont {Y.}~\bibnamefont {Huang}}, \bibinfo {author}
  {\bibfnamefont {Z.}~\bibnamefont {Liu}}, \bibinfo {author} {\bibfnamefont
  {H.}~\bibnamefont {Lei}}, \bibinfo {author} {\bibfnamefont {Z.}~\bibnamefont
  {Yin}},\ and\ \bibinfo {author} {\bibfnamefont {S.}~\bibnamefont {Wang}},\
  }\bibfield  {title} {\bibinfo {title} {{Dirac cone, flat band and saddle
  point in kagome magnet YMn$_6$Sn$_6$}},\ }\href
  {https://doi.org/10.1038/s41467-021-23536-8} {\bibfield  {journal} {\bibinfo
  {journal} {Nature Communications}\ }\textbf {\bibinfo {volume} {12}},\
  \bibinfo {pages} {3129} (\bibinfo {year} {2021})}\BibitemShut {NoStop}%
\bibitem [{\citenamefont {Liu}\ \emph {et~al.}(2020)\citenamefont {Liu},
  \citenamefont {Li}, \citenamefont {Wang}, \citenamefont {Wang}, \citenamefont
  {Wen}, \citenamefont {Jiang}, \citenamefont {Lu}, \citenamefont {Yan},
  \citenamefont {Huang}, \citenamefont {Shen}, \citenamefont {Yin},
  \citenamefont {Wang}, \citenamefont {Yin}, \citenamefont {Lei},\ and\
  \citenamefont {Wang}}]{Liu2020}%
  \BibitemOpen
  \bibfield  {author} {\bibinfo {author} {\bibfnamefont {Z.}~\bibnamefont
  {Liu}}, \bibinfo {author} {\bibfnamefont {M.}~\bibnamefont {Li}}, \bibinfo
  {author} {\bibfnamefont {Q.}~\bibnamefont {Wang}}, \bibinfo {author}
  {\bibfnamefont {G.}~\bibnamefont {Wang}}, \bibinfo {author} {\bibfnamefont
  {C.}~\bibnamefont {Wen}}, \bibinfo {author} {\bibfnamefont {K.}~\bibnamefont
  {Jiang}}, \bibinfo {author} {\bibfnamefont {X.}~\bibnamefont {Lu}}, \bibinfo
  {author} {\bibfnamefont {S.}~\bibnamefont {Yan}}, \bibinfo {author}
  {\bibfnamefont {Y.}~\bibnamefont {Huang}}, \bibinfo {author} {\bibfnamefont
  {D.}~\bibnamefont {Shen}}, \bibinfo {author} {\bibfnamefont {J.-X.}\
  \bibnamefont {Yin}}, \bibinfo {author} {\bibfnamefont {Z.}~\bibnamefont
  {Wang}}, \bibinfo {author} {\bibfnamefont {Z.}~\bibnamefont {Yin}}, \bibinfo
  {author} {\bibfnamefont {H.}~\bibnamefont {Lei}},\ and\ \bibinfo {author}
  {\bibfnamefont {S.}~\bibnamefont {Wang}},\ }\bibfield  {title} {\bibinfo
  {title} {{Orbital-selective Dirac fermions and extremely flat bands in
  frustrated kagome-lattice metal CoSn}},\ }\href
  {https://doi.org/10.1038/s41467-020-17462-4} {\bibfield  {journal} {\bibinfo
  {journal} {Nature Communications}\ }\textbf {\bibinfo {volume} {11}},\
  \bibinfo {pages} {4002} (\bibinfo {year} {2020})}\BibitemShut {NoStop}%
\bibitem [{\citenamefont {Liu}\ \emph {et~al.}(2021)\citenamefont {Liu},
  \citenamefont {Zhao}, \citenamefont {Li}, \citenamefont {Yin}, \citenamefont
  {Wang}, \citenamefont {Liu}, \citenamefont {Shen}, \citenamefont {Huang},
  \citenamefont {Lei}, \citenamefont {Liu},\ and\ \citenamefont
  {Wang}}]{Liu2021}%
  \BibitemOpen
  \bibfield  {author} {\bibinfo {author} {\bibfnamefont {Z.}~\bibnamefont
  {Liu}}, \bibinfo {author} {\bibfnamefont {N.}~\bibnamefont {Zhao}}, \bibinfo
  {author} {\bibfnamefont {M.}~\bibnamefont {Li}}, \bibinfo {author}
  {\bibfnamefont {Q.}~\bibnamefont {Yin}}, \bibinfo {author} {\bibfnamefont
  {Q.}~\bibnamefont {Wang}}, \bibinfo {author} {\bibfnamefont {Z.}~\bibnamefont
  {Liu}}, \bibinfo {author} {\bibfnamefont {D.}~\bibnamefont {Shen}}, \bibinfo
  {author} {\bibfnamefont {Y.}~\bibnamefont {Huang}}, \bibinfo {author}
  {\bibfnamefont {H.}~\bibnamefont {Lei}}, \bibinfo {author} {\bibfnamefont
  {K.}~\bibnamefont {Liu}},\ and\ \bibinfo {author} {\bibfnamefont
  {S.}~\bibnamefont {Wang}},\ }\bibfield  {title} {\bibinfo {title}
  {{Electronic correlation effects in the kagome magnet
  ${\mathrm{GdMn}}_{6}{\mathrm{Sn}}_{6}$}},\ }\href
  {https://doi.org/10.1103/PhysRevB.104.115122} {\bibfield  {journal} {\bibinfo
   {journal} {Phys. Rev. B}\ }\textbf {\bibinfo {volume} {104}},\ \bibinfo
  {pages} {115122} (\bibinfo {year} {2021})}\BibitemShut {NoStop}%
\bibitem [{\citenamefont {{Venturini}}\ \emph {et~al.}(1991)\citenamefont
  {{Venturini}}, \citenamefont {Chafik El~Idrissi},\ and\ \citenamefont
  {Malaman}}]{Venturni1991}%
  \BibitemOpen
  \bibfield  {author} {\bibinfo {author} {\bibfnamefont {G.}~\bibnamefont
  {{Venturini}}}, \bibinfo {author} {\bibfnamefont {B.}~\bibnamefont {Chafik
  El~Idrissi}},\ and\ \bibinfo {author} {\bibfnamefont {B.}~\bibnamefont
  {Malaman}},\ }\bibfield  {title} {\bibinfo {title} {{Magnetic properties of
  RMn$_6$Sn$_6$ (R = Sc, Y, Gd-Tm, Lu) compounds with HfFe$_6$Ge$_6$ type
  structure}},\ }\href {https://doi.org/10.1016/0304-8853(91)90108-M}
  {\bibfield  {journal} {\bibinfo  {journal} {Journal of Magnetism and Magnetic
  Materials}\ }\textbf {\bibinfo {volume} {94}},\ \bibinfo {pages} {35}
  (\bibinfo {year} {1991})}\BibitemShut {NoStop}%
\bibitem [{\citenamefont {Ghimire}\ \emph {et~al.}(2020)\citenamefont
  {Ghimire}, \citenamefont {Dally}, \citenamefont {Poudel}, \citenamefont
  {Jones}, \citenamefont {Michel}, \citenamefont {Magar}, \citenamefont
  {Bleuel}, \citenamefont {McGuire}, \citenamefont {Jiang}, \citenamefont
  {Mitchell}, \citenamefont {Lynn},\ and\ \citenamefont {Mazin}}]{Ghimire2020}%
  \BibitemOpen
  \bibfield  {author} {\bibinfo {author} {\bibfnamefont {N.~J.}\ \bibnamefont
  {Ghimire}}, \bibinfo {author} {\bibfnamefont {R.~L.}\ \bibnamefont {Dally}},
  \bibinfo {author} {\bibfnamefont {L.}~\bibnamefont {Poudel}}, \bibinfo
  {author} {\bibfnamefont {D.~C.}\ \bibnamefont {Jones}}, \bibinfo {author}
  {\bibfnamefont {D.}~\bibnamefont {Michel}}, \bibinfo {author} {\bibfnamefont
  {N.~T.}\ \bibnamefont {Magar}}, \bibinfo {author} {\bibfnamefont
  {M.}~\bibnamefont {Bleuel}}, \bibinfo {author} {\bibfnamefont {M.~A.}\
  \bibnamefont {McGuire}}, \bibinfo {author} {\bibfnamefont {J.~S.}\
  \bibnamefont {Jiang}}, \bibinfo {author} {\bibfnamefont {J.~F.}\ \bibnamefont
  {Mitchell}}, \bibinfo {author} {\bibfnamefont {J.~W.}\ \bibnamefont {Lynn}},\
  and\ \bibinfo {author} {\bibfnamefont {I.~I.}\ \bibnamefont {Mazin}},\
  }\bibfield  {title} {\bibinfo {title} {{Competing magnetic phases and
  fluctuation-driven scalar spin chirality in the kagome metal
  YMn$_6$Sn$_6$}},\ }\href {https://doi.org/10.1126/sciadv.abe2680} {\bibfield
  {journal} {\bibinfo  {journal} {Science Advances}\ }\textbf {\bibinfo
  {volume} {6}},\ \bibinfo {pages} {eabe2680} (\bibinfo {year}
  {2020})}\BibitemShut {NoStop}%
\bibitem [{\citenamefont {{Yin}}\ \emph {et~al.}(2020)\citenamefont {{Yin}},
  \citenamefont {Ma}, \citenamefont {Cochran}, \citenamefont {Xu},
  \citenamefont {Zhang}, \citenamefont {Tien}, \citenamefont {Shumiya},
  \citenamefont {Cheng}, \citenamefont {Jiang}, \citenamefont {Lian},
  \citenamefont {Song}, \citenamefont {Chang}, \citenamefont {Belopolski},
  \citenamefont {Multer}, \citenamefont {Litskevich}, \citenamefont {Cheng},
  \citenamefont {Yang}, \citenamefont {Swidler}, \citenamefont {Zhou},
  \citenamefont {Lin}, \citenamefont {Neupert}, \citenamefont {Wang},
  \citenamefont {Yao}, \citenamefont {Chang}, \citenamefont {Jia},\ and\
  \citenamefont {Zahid~Hasan}}]{Yin2020}%
  \BibitemOpen
  \bibfield  {author} {\bibinfo {author} {\bibfnamefont {J.-X.}\ \bibnamefont
  {{Yin}}}, \bibinfo {author} {\bibfnamefont {W.}~\bibnamefont {Ma}}, \bibinfo
  {author} {\bibfnamefont {T.~A.}\ \bibnamefont {Cochran}}, \bibinfo {author}
  {\bibfnamefont {X.}~\bibnamefont {Xu}}, \bibinfo {author} {\bibfnamefont
  {S.~S.}\ \bibnamefont {Zhang}}, \bibinfo {author} {\bibfnamefont {H.-J.}\
  \bibnamefont {Tien}}, \bibinfo {author} {\bibfnamefont {N.}~\bibnamefont
  {Shumiya}}, \bibinfo {author} {\bibfnamefont {G.}~\bibnamefont {Cheng}},
  \bibinfo {author} {\bibfnamefont {K.}~\bibnamefont {Jiang}}, \bibinfo
  {author} {\bibfnamefont {B.}~\bibnamefont {Lian}}, \bibinfo {author}
  {\bibfnamefont {Z.}~\bibnamefont {Song}}, \bibinfo {author} {\bibfnamefont
  {G.}~\bibnamefont {Chang}}, \bibinfo {author} {\bibfnamefont
  {I.}~\bibnamefont {Belopolski}}, \bibinfo {author} {\bibfnamefont
  {D.}~\bibnamefont {Multer}}, \bibinfo {author} {\bibfnamefont
  {M.}~\bibnamefont {Litskevich}}, \bibinfo {author} {\bibfnamefont {Z.-J.}\
  \bibnamefont {Cheng}}, \bibinfo {author} {\bibfnamefont {X.~P.}\ \bibnamefont
  {Yang}}, \bibinfo {author} {\bibfnamefont {B.}~\bibnamefont {Swidler}},
  \bibinfo {author} {\bibfnamefont {H.}~\bibnamefont {Zhou}}, \bibinfo {author}
  {\bibfnamefont {H.}~\bibnamefont {Lin}}, \bibinfo {author} {\bibfnamefont
  {T.}~\bibnamefont {Neupert}}, \bibinfo {author} {\bibfnamefont
  {Z.}~\bibnamefont {Wang}}, \bibinfo {author} {\bibfnamefont {N.}~\bibnamefont
  {Yao}}, \bibinfo {author} {\bibfnamefont {T.-R.}\ \bibnamefont {Chang}},
  \bibinfo {author} {\bibfnamefont {S.}~\bibnamefont {Jia}},\ and\ \bibinfo
  {author} {\bibfnamefont {M.}~\bibnamefont {Zahid~Hasan}},\ }\bibfield
  {title} {\bibinfo {title} {{Quantum-limit Chern topological magnetism in
  TbMn$_{6}$Sn$_{6}$}},\ }\href {https://doi.org/10.1038/s41586-020-2482-7}
  {\bibfield  {journal} {\bibinfo  {journal} {Nature}\ }\textbf {\bibinfo
  {volume} {583}},\ \bibinfo {pages} {533} (\bibinfo {year}
  {2020})}\BibitemShut {NoStop}%
\bibitem [{\citenamefont {Malaman}\ \emph {et~al.}(1999)\citenamefont
  {Malaman}, \citenamefont {Venturini}, \citenamefont {Welter}, \citenamefont
  {Sanchez}, \citenamefont {Vulliet},\ and\ \citenamefont
  {Ressouche}}]{Malaman1999}%
  \BibitemOpen
  \bibfield  {author} {\bibinfo {author} {\bibfnamefont {B.}~\bibnamefont
  {Malaman}}, \bibinfo {author} {\bibfnamefont {G.}~\bibnamefont {Venturini}},
  \bibinfo {author} {\bibfnamefont {R.}~\bibnamefont {Welter}}, \bibinfo
  {author} {\bibfnamefont {J.}~\bibnamefont {Sanchez}}, \bibinfo {author}
  {\bibfnamefont {P.}~\bibnamefont {Vulliet}},\ and\ \bibinfo {author}
  {\bibfnamefont {E.}~\bibnamefont {Ressouche}},\ }\bibfield  {title} {\bibinfo
  {title} {{Magnetic properties of RMn$_6$Sn$_6$ (R=Gd-Er) compounds from
  neutron diffraction and M\"ossbauer measurements}},\ }\href
  {https://doi.org/https://doi.org/10.1016/S0304-8853(99)00300-5} {\bibfield
  {journal} {\bibinfo  {journal} {Journal of Magnetism and Magnetic Materials}\
  }\textbf {\bibinfo {volume} {202}},\ \bibinfo {pages} {519} (\bibinfo {year}
  {1999})}\BibitemShut {NoStop}%
\bibitem [{\citenamefont {Gu}\ \emph {et~al.}(2022)\citenamefont {Gu},
  \citenamefont {Chen}, \citenamefont {Wei}, \citenamefont {Gao}, \citenamefont
  {Liu}, \citenamefont {Du}, \citenamefont {Pei}, \citenamefont {Zhou},
  \citenamefont {Xu}, \citenamefont {Yin}, \citenamefont {Zhao}, \citenamefont
  {Li}, \citenamefont {Jozwiak}, \citenamefont {Bostwick}, \citenamefont
  {Rotenberg}, \citenamefont {Backes}, \citenamefont {Veiga}, \citenamefont
  {Dhesi}, \citenamefont {Hesjedal}, \citenamefont {van~der Laan},
  \citenamefont {Du}, \citenamefont {Jiang}, \citenamefont {Qi}, \citenamefont
  {Li}, \citenamefont {Shi}, \citenamefont {Liu}, \citenamefont {Chen},\ and\
  \citenamefont {Yang}}]{Gu2022}%
  \BibitemOpen
  \bibfield  {author} {\bibinfo {author} {\bibfnamefont {X.}~\bibnamefont
  {Gu}}, \bibinfo {author} {\bibfnamefont {C.}~\bibnamefont {Chen}}, \bibinfo
  {author} {\bibfnamefont {W.~S.}\ \bibnamefont {Wei}}, \bibinfo {author}
  {\bibfnamefont {L.~L.}\ \bibnamefont {Gao}}, \bibinfo {author} {\bibfnamefont
  {J.~Y.}\ \bibnamefont {Liu}}, \bibinfo {author} {\bibfnamefont
  {X.}~\bibnamefont {Du}}, \bibinfo {author} {\bibfnamefont {D.}~\bibnamefont
  {Pei}}, \bibinfo {author} {\bibfnamefont {J.~S.}\ \bibnamefont {Zhou}},
  \bibinfo {author} {\bibfnamefont {R.~Z.}\ \bibnamefont {Xu}}, \bibinfo
  {author} {\bibfnamefont {Z.~X.}\ \bibnamefont {Yin}}, \bibinfo {author}
  {\bibfnamefont {W.~X.}\ \bibnamefont {Zhao}}, \bibinfo {author}
  {\bibfnamefont {Y.~D.}\ \bibnamefont {Li}}, \bibinfo {author} {\bibfnamefont
  {C.}~\bibnamefont {Jozwiak}}, \bibinfo {author} {\bibfnamefont
  {A.}~\bibnamefont {Bostwick}}, \bibinfo {author} {\bibfnamefont
  {E.}~\bibnamefont {Rotenberg}}, \bibinfo {author} {\bibfnamefont
  {D.}~\bibnamefont {Backes}}, \bibinfo {author} {\bibfnamefont {L.~S.~I.}\
  \bibnamefont {Veiga}}, \bibinfo {author} {\bibfnamefont {S.}~\bibnamefont
  {Dhesi}}, \bibinfo {author} {\bibfnamefont {T.}~\bibnamefont {Hesjedal}},
  \bibinfo {author} {\bibfnamefont {G.}~\bibnamefont {van~der Laan}}, \bibinfo
  {author} {\bibfnamefont {H.~F.}\ \bibnamefont {Du}}, \bibinfo {author}
  {\bibfnamefont {W.~J.}\ \bibnamefont {Jiang}}, \bibinfo {author}
  {\bibfnamefont {Y.~P.}\ \bibnamefont {Qi}}, \bibinfo {author} {\bibfnamefont
  {G.}~\bibnamefont {Li}}, \bibinfo {author} {\bibfnamefont {W.~J.}\
  \bibnamefont {Shi}}, \bibinfo {author} {\bibfnamefont {Z.~K.}\ \bibnamefont
  {Liu}}, \bibinfo {author} {\bibfnamefont {Y.~L.}\ \bibnamefont {Chen}},\ and\
  \bibinfo {author} {\bibfnamefont {L.~X.}\ \bibnamefont {Yang}},\ }\bibfield
  {title} {\bibinfo {title} {{Robust kagome electronic structure in the
  topological quantum magnets $X{\mathrm{Mn}}_{6}{\mathrm{Sn}}_{6}$
  $(X=\mathrm{Dy},\mathrm{Tb},\mathrm{Gd},\mathrm{Y})$}},\ }\href
  {https://doi.org/10.1103/PhysRevB.105.155108} {\bibfield  {journal} {\bibinfo
   {journal} {Phys. Rev. B}\ }\textbf {\bibinfo {volume} {105}},\ \bibinfo
  {pages} {155108} (\bibinfo {year} {2022})}\BibitemShut {NoStop}%
\bibitem [{\citenamefont {Xu}\ \emph {et~al.}(2015)\citenamefont {Xu},
  \citenamefont {Lian},\ and\ \citenamefont {Zhang}}]{Xu2015}%
  \BibitemOpen
  \bibfield  {author} {\bibinfo {author} {\bibfnamefont {G.}~\bibnamefont
  {Xu}}, \bibinfo {author} {\bibfnamefont {B.}~\bibnamefont {Lian}},\ and\
  \bibinfo {author} {\bibfnamefont {S.-C.}\ \bibnamefont {Zhang}},\ }\bibfield
  {title} {\bibinfo {title} {{Intrinsic Quantum Anomalous Hall Effect in the
  Kagome Lattice ${\mathrm{Cs}}_{2}{\mathrm{LiMn}}_{3}{\mathrm{F}}_{12}$}},\
  }\href {https://doi.org/10.1103/PhysRevLett.115.186802} {\bibfield  {journal}
  {\bibinfo  {journal} {Phys. Rev. Lett.}\ }\textbf {\bibinfo {volume} {115}},\
  \bibinfo {pages} {186802} (\bibinfo {year} {2015})}\BibitemShut {NoStop}%
\bibitem [{\citenamefont {Ma}\ \emph {et~al.}(2021{\natexlab{a}})\citenamefont
  {Ma}, \citenamefont {Xu}, \citenamefont {Wang}, \citenamefont {Zhou},
  \citenamefont {Marshall}, \citenamefont {Qu}, \citenamefont {Xie},\ and\
  \citenamefont {Jia}}]{Ma2021}%
  \BibitemOpen
  \bibfield  {author} {\bibinfo {author} {\bibfnamefont {W.}~\bibnamefont
  {Ma}}, \bibinfo {author} {\bibfnamefont {X.}~\bibnamefont {Xu}}, \bibinfo
  {author} {\bibfnamefont {Z.}~\bibnamefont {Wang}}, \bibinfo {author}
  {\bibfnamefont {H.}~\bibnamefont {Zhou}}, \bibinfo {author} {\bibfnamefont
  {M.}~\bibnamefont {Marshall}}, \bibinfo {author} {\bibfnamefont
  {Z.}~\bibnamefont {Qu}}, \bibinfo {author} {\bibfnamefont {W.}~\bibnamefont
  {Xie}},\ and\ \bibinfo {author} {\bibfnamefont {S.}~\bibnamefont {Jia}},\
  }\bibfield  {title} {\bibinfo {title} {{Anomalous Hall effect in the
  distorted kagome magnets (Nd,Sm)${\mathrm{Mn}}_{6}{\mathrm{Sn}}_{6}$}},\
  }\href {https://doi.org/10.1103/PhysRevB.103.235109} {\bibfield  {journal}
  {\bibinfo  {journal} {Phys. Rev. B}\ }\textbf {\bibinfo {volume} {103}},\
  \bibinfo {pages} {235109} (\bibinfo {year} {2021}{\natexlab{a}})}\BibitemShut
  {NoStop}%
\bibitem [{\citenamefont {Xu}\ \emph {et~al.}(2021)\citenamefont {Xu},
  \citenamefont {Heitmann}, \citenamefont {Zhang}, \citenamefont {Xu},\ and\
  \citenamefont {Ke}}]{Xu2021}%
  \BibitemOpen
  \bibfield  {author} {\bibinfo {author} {\bibfnamefont {C.~Q.}\ \bibnamefont
  {Xu}}, \bibinfo {author} {\bibfnamefont {T.~W.}\ \bibnamefont {Heitmann}},
  \bibinfo {author} {\bibfnamefont {H.}~\bibnamefont {Zhang}}, \bibinfo
  {author} {\bibfnamefont {X.}~\bibnamefont {Xu}},\ and\ \bibinfo {author}
  {\bibfnamefont {X.}~\bibnamefont {Ke}},\ }\bibfield  {title} {\bibinfo
  {title} {{Magnetic phase transition, magnetoresistance, and anomalous Hall
  effect in Ga-substituted $\mathrm{Y}{\mathrm{Mn}}_{6}{\mathrm{Sn}}_{6}$ with
  a ferromagnetic kagome lattice}},\ }\href
  {https://doi.org/10.1103/PhysRevB.104.024413} {\bibfield  {journal} {\bibinfo
   {journal} {Phys. Rev. B}\ }\textbf {\bibinfo {volume} {104}},\ \bibinfo
  {pages} {024413} (\bibinfo {year} {2021})}\BibitemShut {NoStop}%
\bibitem [{\citenamefont {Xu}\ \emph {et~al.}(2022)\citenamefont {Xu},
  \citenamefont {Yin}, \citenamefont {Ma}, \citenamefont {Tien}, \citenamefont
  {Qiang}, \citenamefont {Reddy}, \citenamefont {Zhou}, \citenamefont {Shen},
  \citenamefont {Lu}, \citenamefont {Chang}, \citenamefont {Qu},\ and\
  \citenamefont {Jia}}]{Xu2022}%
  \BibitemOpen
  \bibfield  {author} {\bibinfo {author} {\bibfnamefont {X.}~\bibnamefont
  {Xu}}, \bibinfo {author} {\bibfnamefont {J.-X.}\ \bibnamefont {Yin}},
  \bibinfo {author} {\bibfnamefont {W.}~\bibnamefont {Ma}}, \bibinfo {author}
  {\bibfnamefont {H.-J.}\ \bibnamefont {Tien}}, \bibinfo {author}
  {\bibfnamefont {X.-B.}\ \bibnamefont {Qiang}}, \bibinfo {author}
  {\bibfnamefont {P.~V.~S.}\ \bibnamefont {Reddy}}, \bibinfo {author}
  {\bibfnamefont {H.}~\bibnamefont {Zhou}}, \bibinfo {author} {\bibfnamefont
  {J.}~\bibnamefont {Shen}}, \bibinfo {author} {\bibfnamefont {H.-Z.}\
  \bibnamefont {Lu}}, \bibinfo {author} {\bibfnamefont {T.-R.}\ \bibnamefont
  {Chang}}, \bibinfo {author} {\bibfnamefont {Z.}~\bibnamefont {Qu}},\ and\
  \bibinfo {author} {\bibfnamefont {S.}~\bibnamefont {Jia}},\ }\bibfield
  {title} {\bibinfo {title} {{Topological charge-entropy scaling in kagome
  Chern magnet TbMn$_6$Sn$_6$}},\ }\href
  {https://doi.org/10.1038/s41467-022-28796-6} {\bibfield  {journal} {\bibinfo
  {journal} {Nature Communications}\ }\textbf {\bibinfo {volume} {13}},\
  \bibinfo {pages} {1197} (\bibinfo {year} {2022})}\BibitemShut {NoStop}%
\bibitem [{\citenamefont {Gao}\ \emph {et~al.}(2021)\citenamefont {Gao},
  \citenamefont {Shen}, \citenamefont {Wang}, \citenamefont {Shi},
  \citenamefont {Zhao}, \citenamefont {Li}, \citenamefont {Cao}, \citenamefont
  {Pei}, \citenamefont {Ge}, \citenamefont {Li}, \citenamefont {Li},
  \citenamefont {Chen}, \citenamefont {Yan},\ and\ \citenamefont
  {Qi}}]{Gao2021}%
  \BibitemOpen
  \bibfield  {author} {\bibinfo {author} {\bibfnamefont {L.}~\bibnamefont
  {Gao}}, \bibinfo {author} {\bibfnamefont {S.}~\bibnamefont {Shen}}, \bibinfo
  {author} {\bibfnamefont {Q.}~\bibnamefont {Wang}}, \bibinfo {author}
  {\bibfnamefont {W.}~\bibnamefont {Shi}}, \bibinfo {author} {\bibfnamefont
  {Y.}~\bibnamefont {Zhao}}, \bibinfo {author} {\bibfnamefont {C.}~\bibnamefont
  {Li}}, \bibinfo {author} {\bibfnamefont {W.}~\bibnamefont {Cao}}, \bibinfo
  {author} {\bibfnamefont {C.}~\bibnamefont {Pei}}, \bibinfo {author}
  {\bibfnamefont {J.-Y.}\ \bibnamefont {Ge}}, \bibinfo {author} {\bibfnamefont
  {G.}~\bibnamefont {Li}}, \bibinfo {author} {\bibfnamefont {J.}~\bibnamefont
  {Li}}, \bibinfo {author} {\bibfnamefont {Y.}~\bibnamefont {Chen}}, \bibinfo
  {author} {\bibfnamefont {S.}~\bibnamefont {Yan}},\ and\ \bibinfo {author}
  {\bibfnamefont {Y.}~\bibnamefont {Qi}},\ }\bibfield  {title} {\bibinfo
  {title} {{Anomalous Hall effect in ferrimagnetic metal RMn$_6$Sn$_6$ (R = Tb,
  Dy, Ho) with clean Mn kagome lattice}},\ }\href
  {https://doi.org/10.1063/5.0061260} {\bibfield  {journal} {\bibinfo
  {journal} {Applied Physics Letters}\ }\textbf {\bibinfo {volume} {119}},\
  \bibinfo {pages} {092405} (\bibinfo {year} {2021})}\BibitemShut {NoStop}%
\bibitem [{\citenamefont {Asaba}\ \emph {et~al.}(2020)\citenamefont {Asaba},
  \citenamefont {Thomas}, \citenamefont {Curtis}, \citenamefont {Thompson},
  \citenamefont {Bauer},\ and\ \citenamefont {Ronning}}]{Asaba2020}%
  \BibitemOpen
  \bibfield  {author} {\bibinfo {author} {\bibfnamefont {T.}~\bibnamefont
  {Asaba}}, \bibinfo {author} {\bibfnamefont {S.~M.}\ \bibnamefont {Thomas}},
  \bibinfo {author} {\bibfnamefont {M.}~\bibnamefont {Curtis}}, \bibinfo
  {author} {\bibfnamefont {J.~D.}\ \bibnamefont {Thompson}}, \bibinfo {author}
  {\bibfnamefont {E.~D.}\ \bibnamefont {Bauer}},\ and\ \bibinfo {author}
  {\bibfnamefont {F.}~\bibnamefont {Ronning}},\ }\bibfield  {title} {\bibinfo
  {title} {{Anomalous Hall effect in the kagome ferrimagnet
  ${\mathrm{GdMn}}_{6}{\mathrm{Sn}}_{6}$}},\ }\href
  {https://doi.org/10.1103/PhysRevB.101.174415} {\bibfield  {journal} {\bibinfo
   {journal} {Phys. Rev. B}\ }\textbf {\bibinfo {volume} {101}},\ \bibinfo
  {pages} {174415} (\bibinfo {year} {2020})}\BibitemShut {NoStop}%
\bibitem [{\citenamefont {Dhakal}\ \emph {et~al.}(2021)\citenamefont {Dhakal},
  \citenamefont {Cheenicode~Kabeer}, \citenamefont {Pathak}, \citenamefont
  {Kabir}, \citenamefont {Poudel}, \citenamefont {Filippone}, \citenamefont
  {Casey}, \citenamefont {Pradhan~Sakhya}, \citenamefont {Regmi}, \citenamefont
  {Sims}, \citenamefont {Dimitri}, \citenamefont {Manfrinetti}, \citenamefont
  {Gofryk}, \citenamefont {Oppeneer},\ and\ \citenamefont
  {Neupane}}]{Dhakal2021}%
  \BibitemOpen
  \bibfield  {author} {\bibinfo {author} {\bibfnamefont {G.}~\bibnamefont
  {Dhakal}}, \bibinfo {author} {\bibfnamefont {F.}~\bibnamefont
  {Cheenicode~Kabeer}}, \bibinfo {author} {\bibfnamefont {A.~K.}\ \bibnamefont
  {Pathak}}, \bibinfo {author} {\bibfnamefont {F.}~\bibnamefont {Kabir}},
  \bibinfo {author} {\bibfnamefont {N.}~\bibnamefont {Poudel}}, \bibinfo
  {author} {\bibfnamefont {R.}~\bibnamefont {Filippone}}, \bibinfo {author}
  {\bibfnamefont {J.}~\bibnamefont {Casey}}, \bibinfo {author} {\bibfnamefont
  {A.}~\bibnamefont {Pradhan~Sakhya}}, \bibinfo {author} {\bibfnamefont
  {S.}~\bibnamefont {Regmi}}, \bibinfo {author} {\bibfnamefont
  {C.}~\bibnamefont {Sims}}, \bibinfo {author} {\bibfnamefont {K.}~\bibnamefont
  {Dimitri}}, \bibinfo {author} {\bibfnamefont {P.}~\bibnamefont
  {Manfrinetti}}, \bibinfo {author} {\bibfnamefont {K.}~\bibnamefont {Gofryk}},
  \bibinfo {author} {\bibfnamefont {P.~M.}\ \bibnamefont {Oppeneer}},\ and\
  \bibinfo {author} {\bibfnamefont {M.}~\bibnamefont {Neupane}},\ }\bibfield
  {title} {\bibinfo {title} {{Anisotropically large anomalous and topological
  Hall effect in a kagome magnet}},\ }\href
  {https://doi.org/10.1103/PhysRevB.104.L161115} {\bibfield  {journal}
  {\bibinfo  {journal} {Phys. Rev. B}\ }\textbf {\bibinfo {volume} {104}},\
  \bibinfo {pages} {L161115} (\bibinfo {year} {2021})}\BibitemShut {NoStop}%
\bibitem [{\citenamefont {Ma}\ \emph {et~al.}(2021{\natexlab{b}})\citenamefont
  {Ma}, \citenamefont {Xu}, \citenamefont {Yin}, \citenamefont {Yang},
  \citenamefont {Zhou}, \citenamefont {Cheng}, \citenamefont {Huang},
  \citenamefont {Qu}, \citenamefont {Wang}, \citenamefont {Hasan},\ and\
  \citenamefont {Jia}}]{Ma2021a}%
  \BibitemOpen
  \bibfield  {author} {\bibinfo {author} {\bibfnamefont {W.}~\bibnamefont
  {Ma}}, \bibinfo {author} {\bibfnamefont {X.}~\bibnamefont {Xu}}, \bibinfo
  {author} {\bibfnamefont {J.-X.}\ \bibnamefont {Yin}}, \bibinfo {author}
  {\bibfnamefont {H.}~\bibnamefont {Yang}}, \bibinfo {author} {\bibfnamefont
  {H.}~\bibnamefont {Zhou}}, \bibinfo {author} {\bibfnamefont {Z.-J.}\
  \bibnamefont {Cheng}}, \bibinfo {author} {\bibfnamefont {Y.}~\bibnamefont
  {Huang}}, \bibinfo {author} {\bibfnamefont {Z.}~\bibnamefont {Qu}}, \bibinfo
  {author} {\bibfnamefont {F.}~\bibnamefont {Wang}}, \bibinfo {author}
  {\bibfnamefont {M.~Z.}\ \bibnamefont {Hasan}},\ and\ \bibinfo {author}
  {\bibfnamefont {S.}~\bibnamefont {Jia}},\ }\bibfield  {title} {\bibinfo
  {title} {{Rare Earth Engineering in $R{\mathrm{Mn}}_{6}{\mathrm{Sn}}_{6}$
  ($R=\text{Gd}\text{\ensuremath{-}}\text{Tm}$, Lu) Topological Kagome
  Magnets}},\ }\href {https://doi.org/10.1103/PhysRevLett.126.246602}
  {\bibfield  {journal} {\bibinfo  {journal} {Phys. Rev. Lett.}\ }\textbf
  {\bibinfo {volume} {126}},\ \bibinfo {pages} {246602} (\bibinfo {year}
  {2021}{\natexlab{b}})}\BibitemShut {NoStop}%
\bibitem [{\citenamefont {Haldane}(1988)}]{Haldane1988}%
  \BibitemOpen
  \bibfield  {author} {\bibinfo {author} {\bibfnamefont {F.~D.~M.}\
  \bibnamefont {Haldane}},\ }\bibfield  {title} {\bibinfo {title} {{Model for a
  Quantum Hall Effect without Landau Levels: Condensed-Matter Realization of
  the "Parity Anomaly"}},\ }\href {https://doi.org/10.1103/PhysRevLett.61.2015}
  {\bibfield  {journal} {\bibinfo  {journal} {Phys. Rev. Lett.}\ }\textbf
  {\bibinfo {volume} {61}},\ \bibinfo {pages} {2015} (\bibinfo {year}
  {1988})}\BibitemShut {NoStop}%
\bibitem [{\citenamefont {Sims}()}]{Sims2022}%
  \BibitemOpen
  \bibfield  {author} {\bibinfo {author} {\bibfnamefont {C.}~\bibnamefont
  {Sims}},\ }\href@noop {} {\bibinfo {title} {{Existence of Chern Gaps in
  Kagome Magnets RMn$_6$Ge$_6$ (R = Nd, Sm, Tb, Dy, Ho, Er, Yb, Lu)}}},\
  \Eprint {https://arxiv.org/abs/2203.09447} {arXiv:2203.09447} \BibitemShut
  {NoStop}%
\bibitem [{\citenamefont {{El Idrissi}}\ \emph {et~al.}(1991)\citenamefont {{El
  Idrissi}}, \citenamefont {Venturini}, \citenamefont {Malaman},\ and\
  \citenamefont {Fruchart}}]{Idrissi1991}%
  \BibitemOpen
  \bibfield  {author} {\bibinfo {author} {\bibfnamefont {B.}~\bibnamefont {{El
  Idrissi}}}, \bibinfo {author} {\bibfnamefont {G.}~\bibnamefont {Venturini}},
  \bibinfo {author} {\bibfnamefont {B.}~\bibnamefont {Malaman}},\ and\ \bibinfo
  {author} {\bibfnamefont {D.}~\bibnamefont {Fruchart}},\ }\bibfield  {title}
  {\bibinfo {title} {{Magnetic structures of TbMn$_6$Sn$_6$ and HoMn$_6$Sn$_6$
  compounds from neutron diffraction study}},\ }\href
  {https://doi.org/https://doi.org/10.1016/0022-5088(91)90359-C} {\bibfield
  {journal} {\bibinfo  {journal} {Journal of the Less Common Metals}\ }\textbf
  {\bibinfo {volume} {175}},\ \bibinfo {pages} {143} (\bibinfo {year}
  {1991})}\BibitemShut {NoStop}%
\bibitem [{SM()}]{SM}%
  \BibitemOpen
  \href@noop {} {\bibinfo {title} {{See Supplemental Material for experimental
  details, extended data analysis, details on phonon and band structure
  calculations (see, also, references \cite{Fratini2014, pbe96, wien2k,
  Wang2021, Gorbunov2012, Homes1993, Tanner2015, vasp1, vasp2, draxl2006}
  therein).}}}\BibitemShut {Stop}%
\bibitem [{\citenamefont {Ortiz}\ \emph {et~al.}(2019)\citenamefont {Ortiz},
  \citenamefont {Gomes}, \citenamefont {Morey}, \citenamefont {Winiarski},
  \citenamefont {Bordelon}, \citenamefont {Mangum}, \citenamefont {Oswald},
  \citenamefont {Rodriguez-Rivera}, \citenamefont {Neilson}, \citenamefont
  {Wilson}, \citenamefont {Ertekin}, \citenamefont {McQueen},\ and\
  \citenamefont {Toberer}}]{Ortiz2019}%
  \BibitemOpen
  \bibfield  {author} {\bibinfo {author} {\bibfnamefont {B.~R.}\ \bibnamefont
  {Ortiz}}, \bibinfo {author} {\bibfnamefont {L.~C.}\ \bibnamefont {Gomes}},
  \bibinfo {author} {\bibfnamefont {J.~R.}\ \bibnamefont {Morey}}, \bibinfo
  {author} {\bibfnamefont {M.}~\bibnamefont {Winiarski}}, \bibinfo {author}
  {\bibfnamefont {M.}~\bibnamefont {Bordelon}}, \bibinfo {author}
  {\bibfnamefont {J.~S.}\ \bibnamefont {Mangum}}, \bibinfo {author}
  {\bibfnamefont {I.~W.~H.}\ \bibnamefont {Oswald}}, \bibinfo {author}
  {\bibfnamefont {J.~A.}\ \bibnamefont {Rodriguez-Rivera}}, \bibinfo {author}
  {\bibfnamefont {J.~R.}\ \bibnamefont {Neilson}}, \bibinfo {author}
  {\bibfnamefont {S.~D.}\ \bibnamefont {Wilson}}, \bibinfo {author}
  {\bibfnamefont {E.}~\bibnamefont {Ertekin}}, \bibinfo {author} {\bibfnamefont
  {T.~M.}\ \bibnamefont {McQueen}},\ and\ \bibinfo {author} {\bibfnamefont
  {E.~S.}\ \bibnamefont {Toberer}},\ }\bibfield  {title} {\bibinfo {title}
  {{New kagome prototype materials: discovery of
  ${\mathrm{KV}}_{3}{\mathrm{Sb}}_{5},{\mathrm{RbV}}_{3}{\mathrm{Sb}}_{5}$, and
  ${\mathrm{CsV}}_{3}{\mathrm{Sb}}_{5}$}},\ }\href
  {https://doi.org/10.1103/PhysRevMaterials.3.094407} {\bibfield  {journal}
  {\bibinfo  {journal} {Phys. Rev. Materials}\ }\textbf {\bibinfo {volume}
  {3}},\ \bibinfo {pages} {094407} (\bibinfo {year} {2019})}\BibitemShut
  {NoStop}%
\bibitem [{\citenamefont {Neupert}\ \emph {et~al.}(2022)\citenamefont
  {Neupert}, \citenamefont {Denner}, \citenamefont {Yin}, \citenamefont
  {Thomale},\ and\ \citenamefont {Hasan}}]{Neupert2022}%
  \BibitemOpen
  \bibfield  {author} {\bibinfo {author} {\bibfnamefont {T.}~\bibnamefont
  {Neupert}}, \bibinfo {author} {\bibfnamefont {M.~M.}\ \bibnamefont {Denner}},
  \bibinfo {author} {\bibfnamefont {J.-X.}\ \bibnamefont {Yin}}, \bibinfo
  {author} {\bibfnamefont {R.}~\bibnamefont {Thomale}},\ and\ \bibinfo {author}
  {\bibfnamefont {M.~Z.}\ \bibnamefont {Hasan}},\ }\bibfield  {title} {\bibinfo
  {title} {{Charge order and superconductivity in kagome materials}},\ }\href
  {https://doi.org/10.1038/s41567-021-01404-y} {\bibfield  {journal} {\bibinfo
  {journal} {Nature Physics}\ }\textbf {\bibinfo {volume} {18}},\ \bibinfo
  {pages} {137} (\bibinfo {year} {2022})}\BibitemShut {NoStop}%
\bibitem [{\citenamefont {Zhang}\ \emph {et~al.}(2022)\citenamefont {Zhang},
  \citenamefont {Koo}, \citenamefont {Xu}, \citenamefont {Sretenovic},
  \citenamefont {Yan},\ and\ \citenamefont {Ke}}]{Zhang2022}%
  \BibitemOpen
  \bibfield  {author} {\bibinfo {author} {\bibfnamefont {H.}~\bibnamefont
  {Zhang}}, \bibinfo {author} {\bibfnamefont {J.}~\bibnamefont {Koo}}, \bibinfo
  {author} {\bibfnamefont {C.}~\bibnamefont {Xu}}, \bibinfo {author}
  {\bibfnamefont {M.}~\bibnamefont {Sretenovic}}, \bibinfo {author}
  {\bibfnamefont {B.}~\bibnamefont {Yan}},\ and\ \bibinfo {author}
  {\bibfnamefont {X.}~\bibnamefont {Ke}},\ }\bibfield  {title} {\bibinfo
  {title} {{Exchange-biased topological transverse thermoelectric effects in a
  Kagome ferrimagnet}},\ }\href {https://doi.org/10.1038/s41467-022-28733-7}
  {\bibfield  {journal} {\bibinfo  {journal} {Nature Communications}\ }\textbf
  {\bibinfo {volume} {13}},\ \bibinfo {pages} {2041} (\bibinfo {year}
  {2022})}\BibitemShut {NoStop}%
\bibitem [{\citenamefont {Cheng}\ \emph {et~al.}()\citenamefont {Cheng},
  \citenamefont {Belopolski}, \citenamefont {Cochran}, \citenamefont {Tien},
  \citenamefont {Yang}, \citenamefont {Ma}, \citenamefont {Yin}, \citenamefont
  {Zhang}, \citenamefont {Jozwiak}, \citenamefont {Bostwick}, \citenamefont
  {Rotenberg}, \citenamefont {Cheng}, \citenamefont {Hossain}, \citenamefont
  {Zhang}, \citenamefont {Shumiya}, \citenamefont {Multer}, \citenamefont
  {Litskevich}, \citenamefont {Jiang}, \citenamefont {Yao}, \citenamefont
  {Lian}, \citenamefont {Chang}, \citenamefont {Jia}, \citenamefont {Chang},\
  and\ \citenamefont {Hasan}}]{Cheng2022}%
  \BibitemOpen
  \bibfield  {author} {\bibinfo {author} {\bibfnamefont {Z.-J.}\ \bibnamefont
  {Cheng}}, \bibinfo {author} {\bibfnamefont {I.}~\bibnamefont {Belopolski}},
  \bibinfo {author} {\bibfnamefont {T.~A.}\ \bibnamefont {Cochran}}, \bibinfo
  {author} {\bibfnamefont {H.-J.}\ \bibnamefont {Tien}}, \bibinfo {author}
  {\bibfnamefont {X.~P.}\ \bibnamefont {Yang}}, \bibinfo {author}
  {\bibfnamefont {W.}~\bibnamefont {Ma}}, \bibinfo {author} {\bibfnamefont
  {J.-X.}\ \bibnamefont {Yin}}, \bibinfo {author} {\bibfnamefont
  {J.}~\bibnamefont {Zhang}}, \bibinfo {author} {\bibfnamefont
  {C.}~\bibnamefont {Jozwiak}}, \bibinfo {author} {\bibfnamefont
  {A.}~\bibnamefont {Bostwick}}, \bibinfo {author} {\bibfnamefont
  {E.}~\bibnamefont {Rotenberg}}, \bibinfo {author} {\bibfnamefont
  {G.}~\bibnamefont {Cheng}}, \bibinfo {author} {\bibfnamefont {M.~S.}\
  \bibnamefont {Hossain}}, \bibinfo {author} {\bibfnamefont {Q.}~\bibnamefont
  {Zhang}}, \bibinfo {author} {\bibfnamefont {N.}~\bibnamefont {Shumiya}},
  \bibinfo {author} {\bibfnamefont {D.}~\bibnamefont {Multer}}, \bibinfo
  {author} {\bibfnamefont {M.}~\bibnamefont {Litskevich}}, \bibinfo {author}
  {\bibfnamefont {Y.}~\bibnamefont {Jiang}}, \bibinfo {author} {\bibfnamefont
  {N.}~\bibnamefont {Yao}}, \bibinfo {author} {\bibfnamefont {B.}~\bibnamefont
  {Lian}}, \bibinfo {author} {\bibfnamefont {G.}~\bibnamefont {Chang}},
  \bibinfo {author} {\bibfnamefont {S.}~\bibnamefont {Jia}}, \bibinfo {author}
  {\bibfnamefont {T.-R.}\ \bibnamefont {Chang}},\ and\ \bibinfo {author}
  {\bibfnamefont {M.~Z.}\ \bibnamefont {Hasan}},\ }\href@noop {} {\bibinfo
  {title} {{Magnetization-direction-tunable kagome Weyl line}}},\ \Eprint
  {https://arxiv.org/abs/2203.10648} {arXiv:2203.10648} \BibitemShut {NoStop}%
\bibitem [{\citenamefont {{Tabatabai Yazdi}}\ \emph {et~al.}(2011)\citenamefont
  {{Tabatabai Yazdi}}, \citenamefont {Tajabor}, \citenamefont {Behdani},
  \citenamefont {{Rezaee Roknabadi}}, \citenamefont {{Sanavi Khoshnoud}},\ and\
  \citenamefont {Pourarian}}]{Yazdi2011}%
  \BibitemOpen
  \bibfield  {author} {\bibinfo {author} {\bibfnamefont {S.}~\bibnamefont
  {{Tabatabai Yazdi}}}, \bibinfo {author} {\bibfnamefont {N.}~\bibnamefont
  {Tajabor}}, \bibinfo {author} {\bibfnamefont {M.}~\bibnamefont {Behdani}},
  \bibinfo {author} {\bibfnamefont {M.}~\bibnamefont {{Rezaee Roknabadi}}},
  \bibinfo {author} {\bibfnamefont {D.}~\bibnamefont {{Sanavi Khoshnoud}}},\
  and\ \bibinfo {author} {\bibfnamefont {F.}~\bibnamefont {Pourarian}},\
  }\bibfield  {title} {\bibinfo {title} {{Magnetoelastic properties of
  GdMn$_6$Sn$_6$ intermetallic compound}},\ }\href
  {https://doi.org/https://doi.org/10.1016/j.jmmm.2011.03.014} {\bibfield
  {journal} {\bibinfo  {journal} {Journal of Magnetism and Magnetic Materials}\
  }\textbf {\bibinfo {volume} {323}},\ \bibinfo {pages} {2070} (\bibinfo {year}
  {2011})}\BibitemShut {NoStop}%
\bibitem [{\citenamefont {Jones}\ \emph {et~al.}()\citenamefont {Jones},
  \citenamefont {Das}, \citenamefont {Bhandari}, \citenamefont {Liu},
  \citenamefont {Siegfried}, \citenamefont {Ghimire}, \citenamefont {Tsirkin},
  \citenamefont {Mazin},\ and\ \citenamefont {Ghimire}}]{Jones2022}%
  \BibitemOpen
  \bibfield  {author} {\bibinfo {author} {\bibfnamefont {D.~C.}\ \bibnamefont
  {Jones}}, \bibinfo {author} {\bibfnamefont {S.}~\bibnamefont {Das}}, \bibinfo
  {author} {\bibfnamefont {H.}~\bibnamefont {Bhandari}}, \bibinfo {author}
  {\bibfnamefont {X.}~\bibnamefont {Liu}}, \bibinfo {author} {\bibfnamefont
  {P.}~\bibnamefont {Siegfried}}, \bibinfo {author} {\bibfnamefont {M.~P.}\
  \bibnamefont {Ghimire}}, \bibinfo {author} {\bibfnamefont {S.~S.}\
  \bibnamefont {Tsirkin}}, \bibinfo {author} {\bibfnamefont {I.~I.}\
  \bibnamefont {Mazin}},\ and\ \bibinfo {author} {\bibfnamefont {N.~J.}\
  \bibnamefont {Ghimire}},\ }\href@noop {} {\bibinfo {title} {{Origin of spin
  reorientation and intrinsic anomalous Hall effect in the kagome ferrimagnet
  TbMn$_6$Sn$_6$}}},\ \Eprint {https://arxiv.org/abs/2203.17246}
  {arXiv:2203.17246} \BibitemShut {NoStop}%
\bibitem [{\citenamefont {Mielke~III}\ \emph {et~al.}(2022)\citenamefont
  {Mielke~III}, \citenamefont {Ma}, \citenamefont {Pomjakushin}, \citenamefont
  {Zaharko}, \citenamefont {Sturniolo}, \citenamefont {Liu}, \citenamefont
  {Ukleev}, \citenamefont {White}, \citenamefont {Yin}, \citenamefont
  {Tsirkin}, \citenamefont {Larsen}, \citenamefont {Cochran}, \citenamefont
  {Medarde}, \citenamefont {Por\'{e}e}, \citenamefont {Das}, \citenamefont
  {Gupta}, \citenamefont {Wang}, \citenamefont {Chang}, \citenamefont {Wang},
  \citenamefont {Khasanov}, \citenamefont {Neupert}, \citenamefont {Amato},
  \citenamefont {Liborio}, \citenamefont {Jia}, \citenamefont {Hasan},
  \citenamefont {Luetkens},\ and\ \citenamefont {Guguchia}}]{Mielke2022}%
  \BibitemOpen
  \bibfield  {author} {\bibinfo {author} {\bibfnamefont {C.}~\bibnamefont
  {Mielke~III}}, \bibinfo {author} {\bibfnamefont {W.~L.}\ \bibnamefont {Ma}},
  \bibinfo {author} {\bibfnamefont {V.}~\bibnamefont {Pomjakushin}}, \bibinfo
  {author} {\bibfnamefont {O.}~\bibnamefont {Zaharko}}, \bibinfo {author}
  {\bibfnamefont {S.}~\bibnamefont {Sturniolo}}, \bibinfo {author}
  {\bibfnamefont {X.}~\bibnamefont {Liu}}, \bibinfo {author} {\bibfnamefont
  {V.}~\bibnamefont {Ukleev}}, \bibinfo {author} {\bibfnamefont {J.~S.}\
  \bibnamefont {White}}, \bibinfo {author} {\bibfnamefont {J.-X.}\ \bibnamefont
  {Yin}}, \bibinfo {author} {\bibfnamefont {S.~S.}\ \bibnamefont {Tsirkin}},
  \bibinfo {author} {\bibfnamefont {C.~B.}\ \bibnamefont {Larsen}}, \bibinfo
  {author} {\bibfnamefont {T.~A.}\ \bibnamefont {Cochran}}, \bibinfo {author}
  {\bibfnamefont {M.}~\bibnamefont {Medarde}}, \bibinfo {author} {\bibfnamefont
  {V.}~\bibnamefont {Por\'{e}e}}, \bibinfo {author} {\bibfnamefont
  {D.}~\bibnamefont {Das}}, \bibinfo {author} {\bibfnamefont {R.}~\bibnamefont
  {Gupta}}, \bibinfo {author} {\bibfnamefont {C.~N.}\ \bibnamefont {Wang}},
  \bibinfo {author} {\bibfnamefont {J.}~\bibnamefont {Chang}}, \bibinfo
  {author} {\bibfnamefont {Z.~Q.}\ \bibnamefont {Wang}}, \bibinfo {author}
  {\bibfnamefont {R.}~\bibnamefont {Khasanov}}, \bibinfo {author}
  {\bibfnamefont {T.}~\bibnamefont {Neupert}}, \bibinfo {author} {\bibfnamefont
  {A.}~\bibnamefont {Amato}}, \bibinfo {author} {\bibfnamefont
  {L.}~\bibnamefont {Liborio}}, \bibinfo {author} {\bibfnamefont
  {S.}~\bibnamefont {Jia}}, \bibinfo {author} {\bibfnamefont {M.~Z.}\
  \bibnamefont {Hasan}}, \bibinfo {author} {\bibfnamefont {H.}~\bibnamefont
  {Luetkens}},\ and\ \bibinfo {author} {\bibfnamefont {Z.}~\bibnamefont
  {Guguchia}},\ }\bibfield  {title} {\bibinfo {title} {{Low-temperature
  magnetic crossover in the topological kagome magnet TbMn$_6$Sn$_6$}},\ }\href
  {https://doi.org/10.1038/s42005-022-00885-4} {\bibfield  {journal} {\bibinfo
  {journal} {Communications Physics}\ }\textbf {\bibinfo {volume} {5}},\
  \bibinfo {pages} {2399} (\bibinfo {year} {2022})}\BibitemShut {NoStop}%
\bibitem [{\citenamefont {Biswas}\ \emph {et~al.}(2020)\citenamefont {Biswas},
  \citenamefont {Iakutkina}, \citenamefont {Wang}, \citenamefont {Lei},
  \citenamefont {Dressel},\ and\ \citenamefont {Uykur}}]{Biswas2020}%
  \BibitemOpen
  \bibfield  {author} {\bibinfo {author} {\bibfnamefont {A.}~\bibnamefont
  {Biswas}}, \bibinfo {author} {\bibfnamefont {O.}~\bibnamefont {Iakutkina}},
  \bibinfo {author} {\bibfnamefont {Q.}~\bibnamefont {Wang}}, \bibinfo {author}
  {\bibfnamefont {H.~C.}\ \bibnamefont {Lei}}, \bibinfo {author} {\bibfnamefont
  {M.}~\bibnamefont {Dressel}},\ and\ \bibinfo {author} {\bibfnamefont
  {E.}~\bibnamefont {Uykur}},\ }\bibfield  {title} {\bibinfo {title}
  {{Spin-Reorientation-Induced Band Gap in
  ${\mathrm{Fe}}_{3}{\mathrm{Sn}}_{2}$: Optical Signatures of Weyl Nodes}},\
  }\href {https://doi.org/10.1103/PhysRevLett.125.076403} {\bibfield  {journal}
  {\bibinfo  {journal} {Phys. Rev. Lett.}\ }\textbf {\bibinfo {volume} {125}},\
  \bibinfo {pages} {076403} (\bibinfo {year} {2020})}\BibitemShut {NoStop}%
\bibitem [{\citenamefont {Schilberth}\ \emph {et~al.}(2022)\citenamefont
  {Schilberth}, \citenamefont {Unglert}, \citenamefont {Prodan}, \citenamefont
  {Meggle}, \citenamefont {Ebad~Allah}, \citenamefont {Kuntscher},
  \citenamefont {Tsirlin}, \citenamefont {Tsurkan}, \citenamefont
  {Deisenhofer}, \citenamefont {Chioncel}, \citenamefont {K\'ezsm\'arki},\ and\
  \citenamefont {Bord\'acs}}]{Schilberth2021}%
  \BibitemOpen
  \bibfield  {author} {\bibinfo {author} {\bibfnamefont {F.}~\bibnamefont
  {Schilberth}}, \bibinfo {author} {\bibfnamefont {N.}~\bibnamefont {Unglert}},
  \bibinfo {author} {\bibfnamefont {L.}~\bibnamefont {Prodan}}, \bibinfo
  {author} {\bibfnamefont {F.}~\bibnamefont {Meggle}}, \bibinfo {author}
  {\bibfnamefont {J.}~\bibnamefont {Ebad~Allah}}, \bibinfo {author}
  {\bibfnamefont {C.~A.}\ \bibnamefont {Kuntscher}}, \bibinfo {author}
  {\bibfnamefont {A.~A.}\ \bibnamefont {Tsirlin}}, \bibinfo {author}
  {\bibfnamefont {V.}~\bibnamefont {Tsurkan}}, \bibinfo {author} {\bibfnamefont
  {J.}~\bibnamefont {Deisenhofer}}, \bibinfo {author} {\bibfnamefont
  {L.}~\bibnamefont {Chioncel}}, \bibinfo {author} {\bibfnamefont
  {I.}~\bibnamefont {K\'ezsm\'arki}},\ and\ \bibinfo {author} {\bibfnamefont
  {S.}~\bibnamefont {Bord\'acs}},\ }\bibfield  {title} {\bibinfo {title}
  {{Magneto-optical detection of topological contributions to the anomalous
  Hall effect in a kagome ferromagnet}},\ }\href
  {https://doi.org/10.1103/PhysRevB.106.144404} {\bibfield  {journal} {\bibinfo
   {journal} {Phys. Rev. B}\ }\textbf {\bibinfo {volume} {106}},\ \bibinfo
  {pages} {144404} (\bibinfo {year} {2022})}\BibitemShut {NoStop}%
\bibitem [{\citenamefont {Wenzel}\ \emph {et~al.}(2022)\citenamefont {Wenzel},
  \citenamefont {Ortiz}, \citenamefont {Wilson}, \citenamefont {Dressel},
  \citenamefont {Tsirlin},\ and\ \citenamefont {Uykur}}]{Wenzel2022}%
  \BibitemOpen
  \bibfield  {author} {\bibinfo {author} {\bibfnamefont {M.}~\bibnamefont
  {Wenzel}}, \bibinfo {author} {\bibfnamefont {B.~R.}\ \bibnamefont {Ortiz}},
  \bibinfo {author} {\bibfnamefont {S.~D.}\ \bibnamefont {Wilson}}, \bibinfo
  {author} {\bibfnamefont {M.}~\bibnamefont {Dressel}}, \bibinfo {author}
  {\bibfnamefont {A.~A.}\ \bibnamefont {Tsirlin}},\ and\ \bibinfo {author}
  {\bibfnamefont {E.}~\bibnamefont {Uykur}},\ }\bibfield  {title} {\bibinfo
  {title} {{Optical study of ${\mathrm{RbV}}_{3}{\mathrm{Sb}}_{5}$: Multiple
  density-wave gaps and phonon anomalies}},\ }\href
  {https://doi.org/10.1103/PhysRevB.105.245123} {\bibfield  {journal} {\bibinfo
   {journal} {Phys. Rev. B}\ }\textbf {\bibinfo {volume} {105}},\ \bibinfo
  {pages} {245123} (\bibinfo {year} {2022})}\BibitemShut {NoStop}%
\bibitem [{\citenamefont {Uykur}\ \emph {et~al.}(2022)\citenamefont {Uykur},
  \citenamefont {Ortiz}, \citenamefont {Wilson}, \citenamefont {Dressel},\ and\
  \citenamefont {Tsirlin}}]{Uykur2022}%
  \BibitemOpen
  \bibfield  {author} {\bibinfo {author} {\bibfnamefont {E.}~\bibnamefont
  {Uykur}}, \bibinfo {author} {\bibfnamefont {B.~R.}\ \bibnamefont {Ortiz}},
  \bibinfo {author} {\bibfnamefont {S.~D.}\ \bibnamefont {Wilson}}, \bibinfo
  {author} {\bibfnamefont {M.}~\bibnamefont {Dressel}},\ and\ \bibinfo {author}
  {\bibfnamefont {A.~A.}\ \bibnamefont {Tsirlin}},\ }\bibfield  {title}
  {\bibinfo {title} {{Optical detection of the density-wave instability in the
  kagome metal KV$_3$Sb$_5$}},\ }\href
  {https://doi.org/10.1038/s41535-021-00420-8} {\bibfield  {journal} {\bibinfo
  {journal} {npj Quantum Materials}\ }\textbf {\bibinfo {volume} {7}},\
  \bibinfo {pages} {16} (\bibinfo {year} {2022})}\BibitemShut {NoStop}%
\bibitem [{\citenamefont {Uykur}\ \emph {et~al.}(2021)\citenamefont {Uykur},
  \citenamefont {Ortiz}, \citenamefont {Iakutkina}, \citenamefont {Wenzel},
  \citenamefont {Wilson}, \citenamefont {Dressel},\ and\ \citenamefont
  {Tsirlin}}]{Uykur2021}%
  \BibitemOpen
  \bibfield  {author} {\bibinfo {author} {\bibfnamefont {E.}~\bibnamefont
  {Uykur}}, \bibinfo {author} {\bibfnamefont {B.~R.}\ \bibnamefont {Ortiz}},
  \bibinfo {author} {\bibfnamefont {O.}~\bibnamefont {Iakutkina}}, \bibinfo
  {author} {\bibfnamefont {M.}~\bibnamefont {Wenzel}}, \bibinfo {author}
  {\bibfnamefont {S.~D.}\ \bibnamefont {Wilson}}, \bibinfo {author}
  {\bibfnamefont {M.}~\bibnamefont {Dressel}},\ and\ \bibinfo {author}
  {\bibfnamefont {A.~A.}\ \bibnamefont {Tsirlin}},\ }\bibfield  {title}
  {\bibinfo {title} {{Low-energy optical properties of the nonmagnetic kagome
  metal ${\mathrm{CsV}}_{3}{\mathrm{Sb}}_{5}$}},\ }\href
  {https://doi.org/10.1103/PhysRevB.104.045130} {\bibfield  {journal} {\bibinfo
   {journal} {Phys. Rev. B}\ }\textbf {\bibinfo {volume} {104}},\ \bibinfo
  {pages} {045130} (\bibinfo {year} {2021})}\BibitemShut {NoStop}%
\bibitem [{\citenamefont {Kumar}\ \emph {et~al.}(2019)\citenamefont {Kumar},
  \citenamefont {Soh}, \citenamefont {Wang},\ and\ \citenamefont
  {Xiong}}]{Kumar2019}%
  \BibitemOpen
  \bibfield  {author} {\bibinfo {author} {\bibfnamefont {N.}~\bibnamefont
  {Kumar}}, \bibinfo {author} {\bibfnamefont {Y.}~\bibnamefont {Soh}}, \bibinfo
  {author} {\bibfnamefont {Y.}~\bibnamefont {Wang}},\ and\ \bibinfo {author}
  {\bibfnamefont {Y.}~\bibnamefont {Xiong}},\ }\bibfield  {title} {\bibinfo
  {title} {{Magnetotransport as a diagnostic of spin reorientation: Kagome
  ferromagnet as a case study}},\ }\href
  {https://doi.org/10.1103/PhysRevB.100.214420} {\bibfield  {journal} {\bibinfo
   {journal} {Phys. Rev. B}\ }\textbf {\bibinfo {volume} {100}},\ \bibinfo
  {pages} {214420} (\bibinfo {year} {2019})}\BibitemShut {NoStop}%
\bibitem [{\citenamefont {Luo}\ \emph {et~al.}(2022)\citenamefont {Luo},
  \citenamefont {Gao}, \citenamefont {Liu}, \citenamefont {Gu}, \citenamefont
  {Wu}, \citenamefont {Yi}, \citenamefont {Jia}, \citenamefont {Wu},
  \citenamefont {Luo}, \citenamefont {Xu}, \citenamefont {Zhao}, \citenamefont
  {Wang}, \citenamefont {Mao}, \citenamefont {Liu}, \citenamefont {Zhu},
  \citenamefont {Shi}, \citenamefont {Jiang}, \citenamefont {Hu}, \citenamefont
  {Xu},\ and\ \citenamefont {Zhou}}]{Luo2022}%
  \BibitemOpen
  \bibfield  {author} {\bibinfo {author} {\bibfnamefont {H.}~\bibnamefont
  {Luo}}, \bibinfo {author} {\bibfnamefont {Q.}~\bibnamefont {Gao}}, \bibinfo
  {author} {\bibfnamefont {H.}~\bibnamefont {Liu}}, \bibinfo {author}
  {\bibfnamefont {Y.}~\bibnamefont {Gu}}, \bibinfo {author} {\bibfnamefont
  {D.}~\bibnamefont {Wu}}, \bibinfo {author} {\bibfnamefont {C.}~\bibnamefont
  {Yi}}, \bibinfo {author} {\bibfnamefont {J.}~\bibnamefont {Jia}}, \bibinfo
  {author} {\bibfnamefont {S.}~\bibnamefont {Wu}}, \bibinfo {author}
  {\bibfnamefont {X.}~\bibnamefont {Luo}}, \bibinfo {author} {\bibfnamefont
  {Y.}~\bibnamefont {Xu}}, \bibinfo {author} {\bibfnamefont {L.}~\bibnamefont
  {Zhao}}, \bibinfo {author} {\bibfnamefont {Q.}~\bibnamefont {Wang}}, \bibinfo
  {author} {\bibfnamefont {H.}~\bibnamefont {Mao}}, \bibinfo {author}
  {\bibfnamefont {G.}~\bibnamefont {Liu}}, \bibinfo {author} {\bibfnamefont
  {Z.}~\bibnamefont {Zhu}}, \bibinfo {author} {\bibfnamefont {Y.}~\bibnamefont
  {Shi}}, \bibinfo {author} {\bibfnamefont {K.}~\bibnamefont {Jiang}}, \bibinfo
  {author} {\bibfnamefont {J.}~\bibnamefont {Hu}}, \bibinfo {author}
  {\bibfnamefont {Z.}~\bibnamefont {Xu}},\ and\ \bibinfo {author}
  {\bibfnamefont {X.~J.}\ \bibnamefont {Zhou}},\ }\bibfield  {title} {\bibinfo
  {title} {{Electronic nature of charge density wave and electron-phonon
  coupling in kagome superconductor KV$_3$Sb$_5$}},\ }\href
  {https://doi.org/10.1038/s41467-021-27946-6} {\bibfield  {journal} {\bibinfo
  {journal} {Nature Communications}\ }\textbf {\bibinfo {volume} {13}},\
  \bibinfo {pages} {273} (\bibinfo {year} {2022})}\BibitemShut {NoStop}%
\bibitem [{\citenamefont {Zhong}\ \emph {et~al.}()\citenamefont {Zhong},
  \citenamefont {Li}, \citenamefont {Liu}, \citenamefont {Dong}, \citenamefont
  {Arai}, \citenamefont {Li}, \citenamefont {Shi}, \citenamefont {Wang},
  \citenamefont {Shin}, \citenamefont {Lee}, \citenamefont {Miao},
  \citenamefont {Kondo},\ and\ \citenamefont {Okazaki}}]{Zhong2022}%
  \BibitemOpen
  \bibfield  {author} {\bibinfo {author} {\bibfnamefont {Y.}~\bibnamefont
  {Zhong}}, \bibinfo {author} {\bibfnamefont {S.}~\bibnamefont {Li}}, \bibinfo
  {author} {\bibfnamefont {H.}~\bibnamefont {Liu}}, \bibinfo {author}
  {\bibfnamefont {Y.}~\bibnamefont {Dong}}, \bibinfo {author} {\bibfnamefont
  {Y.}~\bibnamefont {Arai}}, \bibinfo {author} {\bibfnamefont {H.}~\bibnamefont
  {Li}}, \bibinfo {author} {\bibfnamefont {Y.}~\bibnamefont {Shi}}, \bibinfo
  {author} {\bibfnamefont {Z.}~\bibnamefont {Wang}}, \bibinfo {author}
  {\bibfnamefont {S.}~\bibnamefont {Shin}}, \bibinfo {author} {\bibfnamefont
  {H.~N.}\ \bibnamefont {Lee}}, \bibinfo {author} {\bibfnamefont
  {H.}~\bibnamefont {Miao}}, \bibinfo {author} {\bibfnamefont {T.}~\bibnamefont
  {Kondo}},\ and\ \bibinfo {author} {\bibfnamefont {K.}~\bibnamefont
  {Okazaki}},\ }\href@noop {} {\bibinfo {title} {{Testing Electron-phonon
  Coupling for the Superconductivity in Kagome Metal $\rm{CsV_3Sb_5}$}}},\
  \Eprint {https://arxiv.org/abs/2207.02407} {arXiv:2207.02407} \BibitemShut
  {NoStop}%
\bibitem [{\citenamefont {Yin}\ \emph {et~al.}(2020)\citenamefont {Yin},
  \citenamefont {Shumiya}, \citenamefont {Mardanya}, \citenamefont {Wang},
  \citenamefont {Zhang}, \citenamefont {Tien}, \citenamefont {Multer},
  \citenamefont {Jiang}, \citenamefont {Cheng}, \citenamefont {Yao},
  \citenamefont {Wu}, \citenamefont {Wu}, \citenamefont {Deng}, \citenamefont
  {Ye}, \citenamefont {He}, \citenamefont {Chang}, \citenamefont {Liu},
  \citenamefont {Jiang}, \citenamefont {Wang}, \citenamefont {Neupert},
  \citenamefont {Agarwal}, \citenamefont {Chang}, \citenamefont {Chu},
  \citenamefont {Lei},\ and\ \citenamefont {Hasan}}]{Yin2020a}%
  \BibitemOpen
  \bibfield  {author} {\bibinfo {author} {\bibfnamefont {J.-X.}\ \bibnamefont
  {Yin}}, \bibinfo {author} {\bibfnamefont {N.}~\bibnamefont {Shumiya}},
  \bibinfo {author} {\bibfnamefont {S.}~\bibnamefont {Mardanya}}, \bibinfo
  {author} {\bibfnamefont {Q.}~\bibnamefont {Wang}}, \bibinfo {author}
  {\bibfnamefont {S.~S.}\ \bibnamefont {Zhang}}, \bibinfo {author}
  {\bibfnamefont {H.-J.}\ \bibnamefont {Tien}}, \bibinfo {author}
  {\bibfnamefont {D.}~\bibnamefont {Multer}}, \bibinfo {author} {\bibfnamefont
  {Y.}~\bibnamefont {Jiang}}, \bibinfo {author} {\bibfnamefont
  {G.}~\bibnamefont {Cheng}}, \bibinfo {author} {\bibfnamefont
  {N.}~\bibnamefont {Yao}}, \bibinfo {author} {\bibfnamefont {S.}~\bibnamefont
  {Wu}}, \bibinfo {author} {\bibfnamefont {D.}~\bibnamefont {Wu}}, \bibinfo
  {author} {\bibfnamefont {L.}~\bibnamefont {Deng}}, \bibinfo {author}
  {\bibfnamefont {Z.}~\bibnamefont {Ye}}, \bibinfo {author} {\bibfnamefont
  {R.}~\bibnamefont {He}}, \bibinfo {author} {\bibfnamefont {G.}~\bibnamefont
  {Chang}}, \bibinfo {author} {\bibfnamefont {Z.}~\bibnamefont {Liu}}, \bibinfo
  {author} {\bibfnamefont {K.}~\bibnamefont {Jiang}}, \bibinfo {author}
  {\bibfnamefont {Z.}~\bibnamefont {Wang}}, \bibinfo {author} {\bibfnamefont
  {T.}~\bibnamefont {Neupert}}, \bibinfo {author} {\bibfnamefont
  {A.}~\bibnamefont {Agarwal}}, \bibinfo {author} {\bibfnamefont {T.-R.}\
  \bibnamefont {Chang}}, \bibinfo {author} {\bibfnamefont {C.-W.}\ \bibnamefont
  {Chu}}, \bibinfo {author} {\bibfnamefont {H.}~\bibnamefont {Lei}},\ and\
  \bibinfo {author} {\bibfnamefont {M.~Z.}\ \bibnamefont {Hasan}},\ }\bibfield
  {title} {\bibinfo {title} {{Fermion–boson many-body interplay in a
  frustrated kagome paramagnet}},\ }\href
  {https://doi.org/10.1038/s41467-020-17464-2} {\bibfield  {journal} {\bibinfo
  {journal} {Nature Communications}\ }\textbf {\bibinfo {volume} {11}},\
  \bibinfo {pages} {273} (\bibinfo {year} {2020})}\BibitemShut {NoStop}%
\bibitem [{\citenamefont {Riberolles}\ \emph {et~al.}(2022)\citenamefont
  {Riberolles}, \citenamefont {Slade}, \citenamefont {Abernathy}, \citenamefont
  {Granroth}, \citenamefont {Li}, \citenamefont {Lee}, \citenamefont
  {Canfield}, \citenamefont {Ueland}, \citenamefont {Ke},\ and\ \citenamefont
  {McQueeney}}]{Riberolles2022}%
  \BibitemOpen
  \bibfield  {author} {\bibinfo {author} {\bibfnamefont {S.~X.~M.}\
  \bibnamefont {Riberolles}}, \bibinfo {author} {\bibfnamefont {T.~J.}\
  \bibnamefont {Slade}}, \bibinfo {author} {\bibfnamefont {D.~L.}\ \bibnamefont
  {Abernathy}}, \bibinfo {author} {\bibfnamefont {G.~E.}\ \bibnamefont
  {Granroth}}, \bibinfo {author} {\bibfnamefont {B.}~\bibnamefont {Li}},
  \bibinfo {author} {\bibfnamefont {Y.}~\bibnamefont {Lee}}, \bibinfo {author}
  {\bibfnamefont {P.~C.}\ \bibnamefont {Canfield}}, \bibinfo {author}
  {\bibfnamefont {B.~G.}\ \bibnamefont {Ueland}}, \bibinfo {author}
  {\bibfnamefont {L.}~\bibnamefont {Ke}},\ and\ \bibinfo {author}
  {\bibfnamefont {R.~J.}\ \bibnamefont {McQueeney}},\ }\bibfield  {title}
  {\bibinfo {title} {{Low-Temperature Competing Magnetic Energy Scales in the
  Topological Ferrimagnet ${\mathrm{TbMn}}_{6}{\mathrm{Sn}}_{6}$}},\ }\href
  {https://doi.org/10.1103/PhysRevX.12.021043} {\bibfield  {journal} {\bibinfo
  {journal} {Phys. Rev. X}\ }\textbf {\bibinfo {volume} {12}},\ \bibinfo
  {pages} {021043} (\bibinfo {year} {2022})}\BibitemShut {NoStop}%
\bibitem [{\citenamefont {Zhang}\ \emph {et~al.}(2020)\citenamefont {Zhang},
  \citenamefont {Feng}, \citenamefont {Heitmann}, \citenamefont {Kolesnikov},
  \citenamefont {Stone}, \citenamefont {Lu},\ and\ \citenamefont
  {Ke}}]{Zhang2020}%
  \BibitemOpen
  \bibfield  {author} {\bibinfo {author} {\bibfnamefont {H.}~\bibnamefont
  {Zhang}}, \bibinfo {author} {\bibfnamefont {X.}~\bibnamefont {Feng}},
  \bibinfo {author} {\bibfnamefont {T.}~\bibnamefont {Heitmann}}, \bibinfo
  {author} {\bibfnamefont {A.~I.}\ \bibnamefont {Kolesnikov}}, \bibinfo
  {author} {\bibfnamefont {M.~B.}\ \bibnamefont {Stone}}, \bibinfo {author}
  {\bibfnamefont {Y.-M.}\ \bibnamefont {Lu}},\ and\ \bibinfo {author}
  {\bibfnamefont {X.}~\bibnamefont {Ke}},\ }\bibfield  {title} {\bibinfo
  {title} {{Topological magnon bands in a room-temperature kagome magnet}},\
  }\href {https://doi.org/10.1103/PhysRevB.101.100405} {\bibfield  {journal}
  {\bibinfo  {journal} {Phys. Rev. B}\ }\textbf {\bibinfo {volume} {101}},\
  \bibinfo {pages} {100405} (\bibinfo {year} {2020})}\BibitemShut {NoStop}%
\bibitem [{\citenamefont {Fratini}\ and\ \citenamefont
  {Ciuchi}(2021)}]{Fratini2021}%
  \BibitemOpen
  \bibfield  {author} {\bibinfo {author} {\bibfnamefont {S.}~\bibnamefont
  {Fratini}}\ and\ \bibinfo {author} {\bibfnamefont {S.}~\bibnamefont
  {Ciuchi}},\ }\bibfield  {title} {\bibinfo {title} {{Displaced Drude peak and
  bad metal from the interaction with slow fluctuations.}},\ }\href
  {https://doi.org/10.21468/SciPostPhys.11.2.039} {\bibfield  {journal}
  {\bibinfo  {journal} {SciPost Phys.}\ }\textbf {\bibinfo {volume} {11}},\
  \bibinfo {pages} {39} (\bibinfo {year} {2021})}\BibitemShut {NoStop}%
\bibitem [{\citenamefont {Delacrétaz}\ \emph {et~al.}(2017)\citenamefont
  {Delacrétaz}, \citenamefont {Goutéraux}, \citenamefont {Hartnoll},\ and\
  \citenamefont {Karlsson}}]{Delacretaz2017}%
  \BibitemOpen
  \bibfield  {author} {\bibinfo {author} {\bibfnamefont {L.~V.}\ \bibnamefont
  {Delacrétaz}}, \bibinfo {author} {\bibfnamefont {B.}~\bibnamefont
  {Goutéraux}}, \bibinfo {author} {\bibfnamefont {S.~A.}\ \bibnamefont
  {Hartnoll}},\ and\ \bibinfo {author} {\bibfnamefont {A.}~\bibnamefont
  {Karlsson}},\ }\bibfield  {title} {\bibinfo {title} {{Bad Metals from
  Fluctuating Density Waves}},\ }\href
  {https://doi.org/10.21468/SciPostPhys.3.3.025} {\bibfield  {journal}
  {\bibinfo  {journal} {SciPost Phys.}\ }\textbf {\bibinfo {volume} {3}},\
  \bibinfo {pages} {025} (\bibinfo {year} {2017})}\BibitemShut {NoStop}%
\bibitem [{\citenamefont {Petersen}\ \emph {et~al.}(2006)\citenamefont
  {Petersen}, \citenamefont {Hafner},\ and\ \citenamefont
  {Marsman}}]{Petersen2006}%
  \BibitemOpen
  \bibfield  {author} {\bibinfo {author} {\bibfnamefont {M.}~\bibnamefont
  {Petersen}}, \bibinfo {author} {\bibfnamefont {J.}~\bibnamefont {Hafner}},\
  and\ \bibinfo {author} {\bibfnamefont {M.}~\bibnamefont {Marsman}},\
  }\bibfield  {title} {\bibinfo {title} {{Structural, electronic and magnetic
  properties of Gd investigated by DFT + $U$ methods: bulk, clean and H-covered
  (0001) surfaces}},\ }\href {https://doi.org/10.1088/0953-8984/18/30/007}
  {\bibfield  {journal} {\bibinfo  {journal} {Journal of Physics: Condensed
  Matter}\ }\textbf {\bibinfo {volume} {18}},\ \bibinfo {pages} {7021}
  (\bibinfo {year} {2006})}\BibitemShut {NoStop}%
\bibitem [{\citenamefont {S\"oderlind}\ \emph {et~al.}(2014)\citenamefont
  {S\"oderlind}, \citenamefont {Turchi}, \citenamefont {Landa},\ and\
  \citenamefont {Lordi}}]{Soderlind2014}%
  \BibitemOpen
  \bibfield  {author} {\bibinfo {author} {\bibfnamefont {P.}~\bibnamefont
  {S\"oderlind}}, \bibinfo {author} {\bibfnamefont {P.~E.~A.}\ \bibnamefont
  {Turchi}}, \bibinfo {author} {\bibfnamefont {A.}~\bibnamefont {Landa}},\ and\
  \bibinfo {author} {\bibfnamefont {V.}~\bibnamefont {Lordi}},\ }\bibfield
  {title} {\bibinfo {title} {{Ground-state properties of rare-earth metals: an
  evaluation of density-functional theory}},\ }\href
  {https://doi.org/10.1088/0953-8984/26/41/416001} {\bibfield  {journal}
  {\bibinfo  {journal} {Journal of Physics: Condensed Matter}\ }\textbf
  {\bibinfo {volume} {26}},\ \bibinfo {pages} {416001} (\bibinfo {year}
  {2014})}\BibitemShut {NoStop}%
\bibitem [{\citenamefont {Lee}\ \emph {et~al.}()\citenamefont {Lee},
  \citenamefont {Skomski}, \citenamefont {Wang}, \citenamefont {Orth},
  \citenamefont {Pathak}, \citenamefont {Harmon}, \citenamefont {McQueeney},
  \citenamefont {Mazin},\ and\ \citenamefont {Ke}}]{Lee2022}%
  \BibitemOpen
  \bibfield  {author} {\bibinfo {author} {\bibfnamefont {Y.}~\bibnamefont
  {Lee}}, \bibinfo {author} {\bibfnamefont {R.}~\bibnamefont {Skomski}},
  \bibinfo {author} {\bibfnamefont {X.}~\bibnamefont {Wang}}, \bibinfo {author}
  {\bibfnamefont {P.~P.}\ \bibnamefont {Orth}}, \bibinfo {author}
  {\bibfnamefont {A.~K.}\ \bibnamefont {Pathak}}, \bibinfo {author}
  {\bibfnamefont {B.~N.}\ \bibnamefont {Harmon}}, \bibinfo {author}
  {\bibfnamefont {R.~J.}\ \bibnamefont {McQueeney}}, \bibinfo {author}
  {\bibfnamefont {I.~I.}\ \bibnamefont {Mazin}},\ and\ \bibinfo {author}
  {\bibfnamefont {L.}~\bibnamefont {Ke}},\ }\href@noop {} {\bibinfo {title}
  {{Interplay between magnetism and band topology in Kagome magnets
  $R$Mn$_6$Sn$_6$}}},\ \Eprint {https://arxiv.org/abs/2201.11265}
  {arXiv:2201.11265} \BibitemShut {NoStop}%
\bibitem [{\citenamefont {Shao}\ \emph {et~al.}(2020)\citenamefont {Shao},
  \citenamefont {Rudenko}, \citenamefont {Hu}, \citenamefont {Sun},
  \citenamefont {Zhu}, \citenamefont {Moon}, \citenamefont {Millis},
  \citenamefont {Yuan}, \citenamefont {Lichtenstein}, \citenamefont {Smirnov},
  \citenamefont {Mao}, \citenamefont {Katsnelson},\ and\ \citenamefont
  {Basov}}]{Shao2020}%
  \BibitemOpen
  \bibfield  {author} {\bibinfo {author} {\bibfnamefont {Y.}~\bibnamefont
  {Shao}}, \bibinfo {author} {\bibfnamefont {A.~N.}\ \bibnamefont {Rudenko}},
  \bibinfo {author} {\bibfnamefont {J.}~\bibnamefont {Hu}}, \bibinfo {author}
  {\bibfnamefont {Z.}~\bibnamefont {Sun}}, \bibinfo {author} {\bibfnamefont
  {Y.}~\bibnamefont {Zhu}}, \bibinfo {author} {\bibfnamefont {S.}~\bibnamefont
  {Moon}}, \bibinfo {author} {\bibfnamefont {A.~J.}\ \bibnamefont {Millis}},
  \bibinfo {author} {\bibfnamefont {S.}~\bibnamefont {Yuan}}, \bibinfo {author}
  {\bibfnamefont {A.~I.}\ \bibnamefont {Lichtenstein}}, \bibinfo {author}
  {\bibfnamefont {D.}~\bibnamefont {Smirnov}}, \bibinfo {author} {\bibfnamefont
  {Z.~Q.}\ \bibnamefont {Mao}}, \bibinfo {author} {\bibfnamefont {M.~I.}\
  \bibnamefont {Katsnelson}},\ and\ \bibinfo {author} {\bibfnamefont {D.~N.}\
  \bibnamefont {Basov}},\ }\bibfield  {title} {\bibinfo {title} {{Electronic
  correlations in nodal-line semimetals}},\ }\href
  {https://doi.org/10.1038/s41567-020-0859-z} {\bibfield  {journal} {\bibinfo
  {journal} {Nature Physics}\ }\textbf {\bibinfo {volume} {16}},\ \bibinfo
  {pages} {273} (\bibinfo {year} {2020})}\BibitemShut {NoStop}%
\bibitem [{\citenamefont {Qazilbash}\ \emph {et~al.}(2009)\citenamefont
  {Qazilbash}, \citenamefont {Hamlin}, \citenamefont {Baumbach}, \citenamefont
  {Zhang}, \citenamefont {Singh}, \citenamefont {Maple},\ and\ \citenamefont
  {Basov}}]{Quazilbash2009}%
  \BibitemOpen
  \bibfield  {author} {\bibinfo {author} {\bibfnamefont {M.~M.}\ \bibnamefont
  {Qazilbash}}, \bibinfo {author} {\bibfnamefont {J.~J.}\ \bibnamefont
  {Hamlin}}, \bibinfo {author} {\bibfnamefont {R.~E.}\ \bibnamefont
  {Baumbach}}, \bibinfo {author} {\bibfnamefont {L.}~\bibnamefont {Zhang}},
  \bibinfo {author} {\bibfnamefont {D.~J.}\ \bibnamefont {Singh}}, \bibinfo
  {author} {\bibfnamefont {M.~B.}\ \bibnamefont {Maple}},\ and\ \bibinfo
  {author} {\bibfnamefont {D.~N.}\ \bibnamefont {Basov}},\ }\bibfield  {title}
  {\bibinfo {title} {{Electronic correlations in the iron pnictides}},\ }\href
  {https://doi.org/10.1038/nphys1343} {\bibfield  {journal} {\bibinfo
  {journal} {Nature Physics}\ }\textbf {\bibinfo {volume} {5}},\ \bibinfo
  {pages} {273} (\bibinfo {year} {2009})}\BibitemShut {NoStop}%
\bibitem [{\citenamefont {Fratini}\ \emph {et~al.}(2014)\citenamefont
  {Fratini}, \citenamefont {Ciuchi},\ and\ \citenamefont
  {Mayou}}]{Fratini2014}%
  \BibitemOpen
  \bibfield  {author} {\bibinfo {author} {\bibfnamefont {S.}~\bibnamefont
  {Fratini}}, \bibinfo {author} {\bibfnamefont {S.}~\bibnamefont {Ciuchi}},\
  and\ \bibinfo {author} {\bibfnamefont {D.}~\bibnamefont {Mayou}},\ }\bibfield
   {title} {\bibinfo {title} {Phenomenological model for charge dynamics and
  optical response of disordered systems: Application to organic
  semiconductors},\ }\href {https://doi.org/10.1103/PhysRevB.89.235201}
  {\bibfield  {journal} {\bibinfo  {journal} {Phys. Rev. B}\ }\textbf {\bibinfo
  {volume} {89}},\ \bibinfo {pages} {235201} (\bibinfo {year}
  {2014})}\BibitemShut {NoStop}%
\bibitem [{\citenamefont {Perdew}\ \emph {et~al.}(1996)\citenamefont {Perdew},
  \citenamefont {Burke},\ and\ \citenamefont {Ernzerhof}}]{pbe96}%
  \BibitemOpen
  \bibfield  {author} {\bibinfo {author} {\bibfnamefont {J.~P.}\ \bibnamefont
  {Perdew}}, \bibinfo {author} {\bibfnamefont {K.}~\bibnamefont {Burke}},\ and\
  \bibinfo {author} {\bibfnamefont {M.}~\bibnamefont {Ernzerhof}},\ }\bibfield
  {title} {\bibinfo {title} {{Generalized Gradient Approximation Made
  Simple}},\ }\href {https://doi.org/10.1103/PhysRevLett.77.3865} {\bibfield
  {journal} {\bibinfo  {journal} {Phys. Rev. Lett.}\ }\textbf {\bibinfo
  {volume} {77}},\ \bibinfo {pages} {3865} (\bibinfo {year}
  {1996})}\BibitemShut {NoStop}%
\bibitem [{\citenamefont {Blaha}\ \emph {et~al.}()\citenamefont {Blaha},
  \citenamefont {Schwarz}, \citenamefont {Madsen}, \citenamefont {Kvasnicka},
  \citenamefont {Luitz}, \citenamefont {Laskowski}, \citenamefont {Tran},\ and\
  \citenamefont {Marks}}]{wien2k}%
  \BibitemOpen
  \bibfield  {author} {\bibinfo {author} {\bibfnamefont {P.}~\bibnamefont
  {Blaha}}, \bibinfo {author} {\bibfnamefont {K.}~\bibnamefont {Schwarz}},
  \bibinfo {author} {\bibfnamefont {G.}~\bibnamefont {Madsen}}, \bibinfo
  {author} {\bibfnamefont {D.}~\bibnamefont {Kvasnicka}}, \bibinfo {author}
  {\bibfnamefont {J.}~\bibnamefont {Luitz}}, \bibinfo {author} {\bibfnamefont
  {R.}~\bibnamefont {Laskowski}}, \bibinfo {author} {\bibfnamefont
  {F.}~\bibnamefont {Tran}},\ and\ \bibinfo {author} {\bibfnamefont
  {L.}~\bibnamefont {Marks}},\ }\href@noop {} {}\bibinfo {note} {WIEN2k, An
  Augmented Plane Wave + Local Orbitals Program for Calculating Crystal
  Properties (Karlheinz Schwarz, Techn. Universit\"at Wien, Austria), 2018.
  ISBN 3-9501031-1-2}\BibitemShut {NoStop}%
\bibitem [{\citenamefont {Wang}\ \emph {et~al.}(2021)\citenamefont {Wang},
  \citenamefont {Neubauer}, \citenamefont {Duan}, \citenamefont {Yin},
  \citenamefont {Fujitsu}, \citenamefont {Hosono}, \citenamefont {Ye},
  \citenamefont {Zhang}, \citenamefont {Chi}, \citenamefont {Krycka},
  \citenamefont {Lei},\ and\ \citenamefont {Dai}}]{Wang2021}%
  \BibitemOpen
  \bibfield  {author} {\bibinfo {author} {\bibfnamefont {Q.}~\bibnamefont
  {Wang}}, \bibinfo {author} {\bibfnamefont {K.~J.}\ \bibnamefont {Neubauer}},
  \bibinfo {author} {\bibfnamefont {C.}~\bibnamefont {Duan}}, \bibinfo {author}
  {\bibfnamefont {Q.}~\bibnamefont {Yin}}, \bibinfo {author} {\bibfnamefont
  {S.}~\bibnamefont {Fujitsu}}, \bibinfo {author} {\bibfnamefont
  {H.}~\bibnamefont {Hosono}}, \bibinfo {author} {\bibfnamefont
  {F.}~\bibnamefont {Ye}}, \bibinfo {author} {\bibfnamefont {R.}~\bibnamefont
  {Zhang}}, \bibinfo {author} {\bibfnamefont {S.}~\bibnamefont {Chi}}, \bibinfo
  {author} {\bibfnamefont {K.}~\bibnamefont {Krycka}}, \bibinfo {author}
  {\bibfnamefont {H.}~\bibnamefont {Lei}},\ and\ \bibinfo {author}
  {\bibfnamefont {P.}~\bibnamefont {Dai}},\ }\bibfield  {title} {\bibinfo
  {title} {{Field-induced topological Hall effect and double-fan spin structure
  with a $c$-axis component in the metallic kagome antiferromagnetic compound
  $\mathrm{Y}{\mathrm{Mn}}_{6}{\mathrm{Sn}}_{6}$}},\ }\href
  {https://doi.org/10.1103/PhysRevB.103.014416} {\bibfield  {journal} {\bibinfo
   {journal} {Phys. Rev. B}\ }\textbf {\bibinfo {volume} {103}},\ \bibinfo
  {pages} {014416} (\bibinfo {year} {2021})}\BibitemShut {NoStop}%
\bibitem [{\citenamefont {Gorbunov}\ \emph {et~al.}(2012)\citenamefont
  {Gorbunov}, \citenamefont {Kuz’min}, \citenamefont {Uhlířová},
  \citenamefont {Žáček}, \citenamefont {Richter}, \citenamefont {Skourski},\
  and\ \citenamefont {Andreev}}]{Gorbunov2012}%
  \BibitemOpen
  \bibfield  {author} {\bibinfo {author} {\bibfnamefont {D.}~\bibnamefont
  {Gorbunov}}, \bibinfo {author} {\bibfnamefont {M.}~\bibnamefont {Kuz’min}},
  \bibinfo {author} {\bibfnamefont {K.}~\bibnamefont {Uhlířová}}, \bibinfo
  {author} {\bibfnamefont {M.}~\bibnamefont {Žáček}}, \bibinfo {author}
  {\bibfnamefont {M.}~\bibnamefont {Richter}}, \bibinfo {author} {\bibfnamefont
  {Y.}~\bibnamefont {Skourski}},\ and\ \bibinfo {author} {\bibfnamefont
  {A.}~\bibnamefont {Andreev}},\ }\bibfield  {title} {\bibinfo {title}
  {Magnetic properties of a gdmn6sn6 single crystal},\ }\href
  {https://doi.org/https://doi.org/10.1016/j.jallcom.2011.12.016} {\bibfield
  {journal} {\bibinfo  {journal} {Journal of Alloys and Compounds}\ }\textbf
  {\bibinfo {volume} {519}},\ \bibinfo {pages} {47} (\bibinfo {year}
  {2012})}\BibitemShut {NoStop}%
\bibitem [{\citenamefont {Homes}\ \emph {et~al.}(1993)\citenamefont {Homes},
  \citenamefont {Reedyk}, \citenamefont {Cradles},\ and\ \citenamefont
  {Timusk}}]{Homes1993}%
  \BibitemOpen
  \bibfield  {author} {\bibinfo {author} {\bibfnamefont {C.~C.}\ \bibnamefont
  {Homes}}, \bibinfo {author} {\bibfnamefont {M.}~\bibnamefont {Reedyk}},
  \bibinfo {author} {\bibfnamefont {D.~A.}\ \bibnamefont {Cradles}},\ and\
  \bibinfo {author} {\bibfnamefont {T.}~\bibnamefont {Timusk}},\ }\bibfield
  {title} {\bibinfo {title} {{Technique for measuring the reflectance of
  irregular, submillimeter-sized samples}},\ }\href
  {https://doi.org/10.1364/AO.32.002976} {\bibfield  {journal} {\bibinfo
  {journal} {Appl. Opt.}\ }\textbf {\bibinfo {volume} {32}},\ \bibinfo {pages}
  {2976} (\bibinfo {year} {1993})}\BibitemShut {NoStop}%
\bibitem [{\citenamefont {Tanner}(2015)}]{Tanner2015}%
  \BibitemOpen
  \bibfield  {author} {\bibinfo {author} {\bibfnamefont {D.~B.}\ \bibnamefont
  {Tanner}},\ }\bibfield  {title} {\bibinfo {title} {{Use of x-ray scattering
  functions in Kramers-Kronig analysis of reflectance}},\ }\href
  {https://doi.org/10.1103/PhysRevB.91.035123} {\bibfield  {journal} {\bibinfo
  {journal} {Phys. Rev. B}\ }\textbf {\bibinfo {volume} {91}},\ \bibinfo
  {pages} {035123} (\bibinfo {year} {2015})}\BibitemShut {NoStop}%
\bibitem [{\citenamefont {Kresse}\ and\ \citenamefont
  {Furthm\"uller}(1996{\natexlab{a}})}]{vasp1}%
  \BibitemOpen
  \bibfield  {author} {\bibinfo {author} {\bibfnamefont {G.}~\bibnamefont
  {Kresse}}\ and\ \bibinfo {author} {\bibfnamefont {J.}~\bibnamefont
  {Furthm\"uller}},\ }\bibfield  {title} {\bibinfo {title} {Efficiency of
  \textit{ab-initio} total energy calculations for metals and semiconductors
  using a plane-wave basis set},\ }\href
  {https://doi.org/10.1016/0927-0256(96)00008-0} {\bibfield  {journal}
  {\bibinfo  {journal} {Computational Materials Science}\ }\textbf {\bibinfo
  {volume} {6}},\ \bibinfo {pages} {15} (\bibinfo {year}
  {1996}{\natexlab{a}})}\BibitemShut {NoStop}%
\bibitem [{\citenamefont {Kresse}\ and\ \citenamefont
  {Furthm\"uller}(1996{\natexlab{b}})}]{vasp2}%
  \BibitemOpen
  \bibfield  {author} {\bibinfo {author} {\bibfnamefont {G.}~\bibnamefont
  {Kresse}}\ and\ \bibinfo {author} {\bibfnamefont {J.}~\bibnamefont
  {Furthm\"uller}},\ }\bibfield  {title} {\bibinfo {title} {Efficient iterative
  schemes for \textit{ab initio} total-energy calculations using a plane-wave
  basis set},\ }\href {https://doi.org/10.1103/PhysRevB.54.11169} {\bibfield
  {journal} {\bibinfo  {journal} {Phys. Rev. B}\ }\textbf {\bibinfo {volume}
  {54}},\ \bibinfo {pages} {11169} (\bibinfo {year}
  {1996}{\natexlab{b}})}\BibitemShut {NoStop}%
\bibitem [{\citenamefont {Ambrosch-Draxl}\ and\ \citenamefont
  {Sofo}(2006)}]{draxl2006}%
  \BibitemOpen
  \bibfield  {author} {\bibinfo {author} {\bibfnamefont {C.}~\bibnamefont
  {Ambrosch-Draxl}}\ and\ \bibinfo {author} {\bibfnamefont {J.}~\bibnamefont
  {Sofo}},\ }\bibfield  {title} {\bibinfo {title} {Linear optical properties of
  solids within the full-potential linearized augmented planewave method},\
  }\href {https://doi.org/10.1016/j.cpc.2006.03.005} {\bibfield  {journal}
  {\bibinfo  {journal} {Comput. Phys. Commun.}\ }\textbf {\bibinfo {volume}
  {175}},\ \bibinfo {pages} {1} (\bibinfo {year} {2006})}\BibitemShut {NoStop}%
\end{thebibliography}%


\begin{thebibliography}{13}%
\makeatletter
\providecommand \@ifxundefined [1]{%
 \@ifx{#1\undefined}
}%
\providecommand \@ifnum [1]{%
 \ifnum #1\expandafter \@firstoftwo
 \else \expandafter \@secondoftwo
 \fi
}%
\providecommand \@ifx [1]{%
 \ifx #1\expandafter \@firstoftwo
 \else \expandafter \@secondoftwo
 \fi
}%
\providecommand \natexlab [1]{#1}%
\providecommand \enquote  [1]{``#1''}%
\providecommand \bibnamefont  [1]{#1}%
\providecommand \bibfnamefont [1]{#1}%
\providecommand \citenamefont [1]{#1}%
\providecommand \href@noop [0]{\@secondoftwo}%
\providecommand \href [0]{\begingroup \@sanitize@url \@href}%
\providecommand \@href[1]{\@@startlink{#1}\@@href}%
\providecommand \@@href[1]{\endgroup#1\@@endlink}%
\providecommand \@sanitize@url [0]{\catcode `\\12\catcode `\$12\catcode
  `\&12\catcode `\#12\catcode `\^12\catcode `\_12\catcode `\%12\relax}%
\providecommand \@@startlink[1]{}%
\providecommand \@@endlink[0]{}%
\providecommand \url  [0]{\begingroup\@sanitize@url \@url }%
\providecommand \@url [1]{\endgroup\@href {#1}{\urlprefix }}%
\providecommand \urlprefix  [0]{URL }%
\providecommand \Eprint [0]{\href }%
\providecommand \doibase [0]{https://doi.org/}%
\providecommand \selectlanguage [0]{\@gobble}%
\providecommand \bibinfo  [0]{\@secondoftwo}%
\providecommand \bibfield  [0]{\@secondoftwo}%
\providecommand \translation [1]{[#1]}%
\providecommand \BibitemOpen [0]{}%
\providecommand \bibitemStop [0]{}%
\providecommand \bibitemNoStop [0]{.\EOS\space}%
\providecommand \EOS [0]{\spacefactor3000\relax}%
\providecommand \BibitemShut  [1]{\csname bibitem#1\endcsname}%
\let\auto@bib@innerbib\@empty
\bibitem [{\citenamefont {Jones}\ \emph {et~al.}()\citenamefont {Jones},
  \citenamefont {Das}, \citenamefont {Bhandari}, \citenamefont {Liu},
  \citenamefont {Siegfried}, \citenamefont {Ghimire}, \citenamefont {Tsirkin},
  \citenamefont {Mazin},\ and\ \citenamefont {Ghimire}}]{Jones2022}%
  \BibitemOpen
  \bibfield  {author} {\bibinfo {author} {\bibfnamefont {D.~C.}\ \bibnamefont
  {Jones}}, \bibinfo {author} {\bibfnamefont {S.}~\bibnamefont {Das}}, \bibinfo
  {author} {\bibfnamefont {H.}~\bibnamefont {Bhandari}}, \bibinfo {author}
  {\bibfnamefont {X.}~\bibnamefont {Liu}}, \bibinfo {author} {\bibfnamefont
  {P.}~\bibnamefont {Siegfried}}, \bibinfo {author} {\bibfnamefont {M.~P.}\
  \bibnamefont {Ghimire}}, \bibinfo {author} {\bibfnamefont {S.~S.}\
  \bibnamefont {Tsirkin}}, \bibinfo {author} {\bibfnamefont {I.~I.}\
  \bibnamefont {Mazin}},\ and\ \bibinfo {author} {\bibfnamefont {N.~J.}\
  \bibnamefont {Ghimire}},\ }\href@noop {} {\bibinfo {title} {{Origin of spin
  reorientation and intrinsic anomalous Hall effect in the kagome ferrimagnet
  TbMn$_6$Sn$_6$}}},\ \Eprint {https://arxiv.org/abs/2203.17246}
  {arXiv:2203.17246} \BibitemShut {NoStop}%
\bibitem [{\citenamefont {Gorbunov}\ \emph {et~al.}(2012)\citenamefont
  {Gorbunov}, \citenamefont {Kuz’min}, \citenamefont {Uhlířová},
  \citenamefont {Žáček}, \citenamefont {Richter}, \citenamefont {Skourski},\
  and\ \citenamefont {Andreev}}]{Gorbunov2012}%
  \BibitemOpen
  \bibfield  {author} {\bibinfo {author} {\bibfnamefont {D.}~\bibnamefont
  {Gorbunov}}, \bibinfo {author} {\bibfnamefont {M.}~\bibnamefont {Kuz’min}},
  \bibinfo {author} {\bibfnamefont {K.}~\bibnamefont {Uhlířová}}, \bibinfo
  {author} {\bibfnamefont {M.}~\bibnamefont {Žáček}}, \bibinfo {author}
  {\bibfnamefont {M.}~\bibnamefont {Richter}}, \bibinfo {author} {\bibfnamefont
  {Y.}~\bibnamefont {Skourski}},\ and\ \bibinfo {author} {\bibfnamefont
  {A.}~\bibnamefont {Andreev}},\ }\bibfield  {title} {\bibinfo {title}
  {Magnetic properties of a gdmn6sn6 single crystal},\ }\href
  {https://doi.org/https://doi.org/10.1016/j.jallcom.2011.12.016} {\bibfield
  {journal} {\bibinfo  {journal} {Journal of Alloys and Compounds}\ }\textbf
  {\bibinfo {volume} {519}},\ \bibinfo {pages} {47} (\bibinfo {year}
  {2012})}\BibitemShut {NoStop}%
\bibitem [{\citenamefont {Homes}\ \emph {et~al.}(1993)\citenamefont {Homes},
  \citenamefont {Reedyk}, \citenamefont {Cradles},\ and\ \citenamefont
  {Timusk}}]{Homes1993}%
  \BibitemOpen
  \bibfield  {author} {\bibinfo {author} {\bibfnamefont {C.~C.}\ \bibnamefont
  {Homes}}, \bibinfo {author} {\bibfnamefont {M.}~\bibnamefont {Reedyk}},
  \bibinfo {author} {\bibfnamefont {D.~A.}\ \bibnamefont {Cradles}},\ and\
  \bibinfo {author} {\bibfnamefont {T.}~\bibnamefont {Timusk}},\ }\bibfield
  {title} {\bibinfo {title} {{Technique for measuring the reflectance of
  irregular, submillimeter-sized samples}},\ }\href
  {https://doi.org/10.1364/AO.32.002976} {\bibfield  {journal} {\bibinfo
  {journal} {Appl. Opt.}\ }\textbf {\bibinfo {volume} {32}},\ \bibinfo {pages}
  {2976} (\bibinfo {year} {1993})}\BibitemShut {NoStop}%
\bibitem [{\citenamefont {Tanner}(2015)}]{Tanner2015}%
  \BibitemOpen
  \bibfield  {author} {\bibinfo {author} {\bibfnamefont {D.~B.}\ \bibnamefont
  {Tanner}},\ }\bibfield  {title} {\bibinfo {title} {{Use of x-ray scattering
  functions in Kramers-Kronig analysis of reflectance}},\ }\href
  {https://doi.org/10.1103/PhysRevB.91.035123} {\bibfield  {journal} {\bibinfo
  {journal} {Phys. Rev. B}\ }\textbf {\bibinfo {volume} {91}},\ \bibinfo
  {pages} {035123} (\bibinfo {year} {2015})}\BibitemShut {NoStop}%
\bibitem [{\citenamefont {Fratini}\ \emph {et~al.}(2014)\citenamefont
  {Fratini}, \citenamefont {Ciuchi},\ and\ \citenamefont
  {Mayou}}]{Fratini2014}%
  \BibitemOpen
  \bibfield  {author} {\bibinfo {author} {\bibfnamefont {S.}~\bibnamefont
  {Fratini}}, \bibinfo {author} {\bibfnamefont {S.}~\bibnamefont {Ciuchi}},\
  and\ \bibinfo {author} {\bibfnamefont {D.}~\bibnamefont {Mayou}},\ }\bibfield
   {title} {\bibinfo {title} {Phenomenological model for charge dynamics and
  optical response of disordered systems: Application to organic
  semiconductors},\ }\href {https://doi.org/10.1103/PhysRevB.89.235201}
  {\bibfield  {journal} {\bibinfo  {journal} {Phys. Rev. B}\ }\textbf {\bibinfo
  {volume} {89}},\ \bibinfo {pages} {235201} (\bibinfo {year}
  {2014})}\BibitemShut {NoStop}%
\bibitem [{\citenamefont {Kresse}\ and\ \citenamefont
  {Furthm\"uller}(1996{\natexlab{a}})}]{vasp1}%
  \BibitemOpen
  \bibfield  {author} {\bibinfo {author} {\bibfnamefont {G.}~\bibnamefont
  {Kresse}}\ and\ \bibinfo {author} {\bibfnamefont {J.}~\bibnamefont
  {Furthm\"uller}},\ }\bibfield  {title} {\bibinfo {title} {Efficiency of
  \textit{ab-initio} total energy calculations for metals and semiconductors
  using a plane-wave basis set},\ }\href
  {https://doi.org/10.1016/0927-0256(96)00008-0} {\bibfield  {journal}
  {\bibinfo  {journal} {Computational Materials Science}\ }\textbf {\bibinfo
  {volume} {6}},\ \bibinfo {pages} {15} (\bibinfo {year}
  {1996}{\natexlab{a}})}\BibitemShut {NoStop}%
\bibitem [{\citenamefont {Kresse}\ and\ \citenamefont
  {Furthm\"uller}(1996{\natexlab{b}})}]{vasp2}%
  \BibitemOpen
  \bibfield  {author} {\bibinfo {author} {\bibfnamefont {G.}~\bibnamefont
  {Kresse}}\ and\ \bibinfo {author} {\bibfnamefont {J.}~\bibnamefont
  {Furthm\"uller}},\ }\bibfield  {title} {\bibinfo {title} {Efficient iterative
  schemes for \textit{ab initio} total-energy calculations using a plane-wave
  basis set},\ }\href {https://doi.org/10.1103/PhysRevB.54.11169} {\bibfield
  {journal} {\bibinfo  {journal} {Phys. Rev. B}\ }\textbf {\bibinfo {volume}
  {54}},\ \bibinfo {pages} {11169} (\bibinfo {year}
  {1996}{\natexlab{b}})}\BibitemShut {NoStop}%
\bibitem [{\citenamefont {Perdew}\ \emph {et~al.}(1996)\citenamefont {Perdew},
  \citenamefont {Burke},\ and\ \citenamefont {Ernzerhof}}]{pbe96}%
  \BibitemOpen
  \bibfield  {author} {\bibinfo {author} {\bibfnamefont {J.~P.}\ \bibnamefont
  {Perdew}}, \bibinfo {author} {\bibfnamefont {K.}~\bibnamefont {Burke}},\ and\
  \bibinfo {author} {\bibfnamefont {M.}~\bibnamefont {Ernzerhof}},\ }\bibfield
  {title} {\bibinfo {title} {{Generalized Gradient Approximation Made
  Simple}},\ }\href {https://doi.org/10.1103/PhysRevLett.77.3865} {\bibfield
  {journal} {\bibinfo  {journal} {Phys. Rev. Lett.}\ }\textbf {\bibinfo
  {volume} {77}},\ \bibinfo {pages} {3865} (\bibinfo {year}
  {1996})}\BibitemShut {NoStop}%
\bibitem [{\citenamefont {Blaha}\ \emph {et~al.}()\citenamefont {Blaha},
  \citenamefont {Schwarz}, \citenamefont {Madsen}, \citenamefont {Kvasnicka},
  \citenamefont {Luitz}, \citenamefont {Laskowski}, \citenamefont {Tran},\ and\
  \citenamefont {Marks}}]{wien2k}%
  \BibitemOpen
  \bibfield  {author} {\bibinfo {author} {\bibfnamefont {P.}~\bibnamefont
  {Blaha}}, \bibinfo {author} {\bibfnamefont {K.}~\bibnamefont {Schwarz}},
  \bibinfo {author} {\bibfnamefont {G.}~\bibnamefont {Madsen}}, \bibinfo
  {author} {\bibfnamefont {D.}~\bibnamefont {Kvasnicka}}, \bibinfo {author}
  {\bibfnamefont {J.}~\bibnamefont {Luitz}}, \bibinfo {author} {\bibfnamefont
  {R.}~\bibnamefont {Laskowski}}, \bibinfo {author} {\bibfnamefont
  {F.}~\bibnamefont {Tran}},\ and\ \bibinfo {author} {\bibfnamefont
  {L.}~\bibnamefont {Marks}},\ }\href@noop {} {}\bibinfo {note} {WIEN2k, An
  Augmented Plane Wave + Local Orbitals Program for Calculating Crystal
  Properties (Karlheinz Schwarz, Techn. Universit\"at Wien, Austria), 2018.
  ISBN 3-9501031-1-2}\BibitemShut {NoStop}%
\bibitem [{\citenamefont {Ambrosch-Draxl}\ and\ \citenamefont
  {Sofo}(2006)}]{draxl2006}%
  \BibitemOpen
  \bibfield  {author} {\bibinfo {author} {\bibfnamefont {C.}~\bibnamefont
  {Ambrosch-Draxl}}\ and\ \bibinfo {author} {\bibfnamefont {J.}~\bibnamefont
  {Sofo}},\ }\bibfield  {title} {\bibinfo {title} {Linear optical properties of
  solids within the full-potential linearized augmented planewave method},\
  }\href {https://doi.org/10.1016/j.cpc.2006.03.005} {\bibfield  {journal}
  {\bibinfo  {journal} {Comput. Phys. Commun.}\ }\textbf {\bibinfo {volume}
  {175}},\ \bibinfo {pages} {1} (\bibinfo {year} {2006})}\BibitemShut {NoStop}%
\bibitem [{\citenamefont {Liu}\ \emph {et~al.}(2021)\citenamefont {Liu},
  \citenamefont {Zhao}, \citenamefont {Li}, \citenamefont {Yin}, \citenamefont
  {Wang}, \citenamefont {Liu}, \citenamefont {Shen}, \citenamefont {Huang},
  \citenamefont {Lei}, \citenamefont {Liu},\ and\ \citenamefont
  {Wang}}]{Liu2021}%
  \BibitemOpen
  \bibfield  {author} {\bibinfo {author} {\bibfnamefont {Z.}~\bibnamefont
  {Liu}}, \bibinfo {author} {\bibfnamefont {N.}~\bibnamefont {Zhao}}, \bibinfo
  {author} {\bibfnamefont {M.}~\bibnamefont {Li}}, \bibinfo {author}
  {\bibfnamefont {Q.}~\bibnamefont {Yin}}, \bibinfo {author} {\bibfnamefont
  {Q.}~\bibnamefont {Wang}}, \bibinfo {author} {\bibfnamefont {Z.}~\bibnamefont
  {Liu}}, \bibinfo {author} {\bibfnamefont {D.}~\bibnamefont {Shen}}, \bibinfo
  {author} {\bibfnamefont {Y.}~\bibnamefont {Huang}}, \bibinfo {author}
  {\bibfnamefont {H.}~\bibnamefont {Lei}}, \bibinfo {author} {\bibfnamefont
  {K.}~\bibnamefont {Liu}},\ and\ \bibinfo {author} {\bibfnamefont
  {S.}~\bibnamefont {Wang}},\ }\bibfield  {title} {\bibinfo {title}
  {{Electronic correlation effects in the kagome magnet
  ${\mathrm{GdMn}}_{6}{\mathrm{Sn}}_{6}$}},\ }\href
  {https://doi.org/10.1103/PhysRevB.104.115122} {\bibfield  {journal} {\bibinfo
   {journal} {Phys. Rev. B}\ }\textbf {\bibinfo {volume} {104}},\ \bibinfo
  {pages} {115122} (\bibinfo {year} {2021})}\BibitemShut {NoStop}%
\bibitem [{\citenamefont {Ma}\ \emph {et~al.}(2021)\citenamefont {Ma},
  \citenamefont {Xu}, \citenamefont {Yin}, \citenamefont {Yang}, \citenamefont
  {Zhou}, \citenamefont {Cheng}, \citenamefont {Huang}, \citenamefont {Qu},
  \citenamefont {Wang}, \citenamefont {Hasan},\ and\ \citenamefont
  {Jia}}]{Ma2021a}%
  \BibitemOpen
  \bibfield  {author} {\bibinfo {author} {\bibfnamefont {W.}~\bibnamefont
  {Ma}}, \bibinfo {author} {\bibfnamefont {X.}~\bibnamefont {Xu}}, \bibinfo
  {author} {\bibfnamefont {J.-X.}\ \bibnamefont {Yin}}, \bibinfo {author}
  {\bibfnamefont {H.}~\bibnamefont {Yang}}, \bibinfo {author} {\bibfnamefont
  {H.}~\bibnamefont {Zhou}}, \bibinfo {author} {\bibfnamefont {Z.-J.}\
  \bibnamefont {Cheng}}, \bibinfo {author} {\bibfnamefont {Y.}~\bibnamefont
  {Huang}}, \bibinfo {author} {\bibfnamefont {Z.}~\bibnamefont {Qu}}, \bibinfo
  {author} {\bibfnamefont {F.}~\bibnamefont {Wang}}, \bibinfo {author}
  {\bibfnamefont {M.~Z.}\ \bibnamefont {Hasan}},\ and\ \bibinfo {author}
  {\bibfnamefont {S.}~\bibnamefont {Jia}},\ }\bibfield  {title} {\bibinfo
  {title} {{Rare Earth Engineering in $R{\mathrm{Mn}}_{6}{\mathrm{Sn}}_{6}$
  ($R=\text{Gd}\text{\ensuremath{-}}\text{Tm}$, Lu) Topological Kagome
  Magnets}},\ }\href {https://doi.org/10.1103/PhysRevLett.126.246602}
  {\bibfield  {journal} {\bibinfo  {journal} {Phys. Rev. Lett.}\ }\textbf
  {\bibinfo {volume} {126}},\ \bibinfo {pages} {246602} (\bibinfo {year}
  {2021})}\BibitemShut {NoStop}%
\bibitem [{\citenamefont {{Yin}}\ \emph {et~al.}(2020)\citenamefont {{Yin}},
  \citenamefont {Ma}, \citenamefont {Cochran}, \citenamefont {Xu},
  \citenamefont {Zhang}, \citenamefont {Tien}, \citenamefont {Shumiya},
  \citenamefont {Cheng}, \citenamefont {Jiang}, \citenamefont {Lian},
  \citenamefont {Song}, \citenamefont {Chang}, \citenamefont {Belopolski},
  \citenamefont {Multer}, \citenamefont {Litskevich}, \citenamefont {Cheng},
  \citenamefont {Yang}, \citenamefont {Swidler}, \citenamefont {Zhou},
  \citenamefont {Lin}, \citenamefont {Neupert}, \citenamefont {Wang},
  \citenamefont {Yao}, \citenamefont {Chang}, \citenamefont {Jia},\ and\
  \citenamefont {Zahid~Hasan}}]{Yin2020}%
  \BibitemOpen
  \bibfield  {author} {\bibinfo {author} {\bibfnamefont {J.-X.}\ \bibnamefont
  {{Yin}}}, \bibinfo {author} {\bibfnamefont {W.}~\bibnamefont {Ma}}, \bibinfo
  {author} {\bibfnamefont {T.~A.}\ \bibnamefont {Cochran}}, \bibinfo {author}
  {\bibfnamefont {X.}~\bibnamefont {Xu}}, \bibinfo {author} {\bibfnamefont
  {S.~S.}\ \bibnamefont {Zhang}}, \bibinfo {author} {\bibfnamefont {H.-J.}\
  \bibnamefont {Tien}}, \bibinfo {author} {\bibfnamefont {N.}~\bibnamefont
  {Shumiya}}, \bibinfo {author} {\bibfnamefont {G.}~\bibnamefont {Cheng}},
  \bibinfo {author} {\bibfnamefont {K.}~\bibnamefont {Jiang}}, \bibinfo
  {author} {\bibfnamefont {B.}~\bibnamefont {Lian}}, \bibinfo {author}
  {\bibfnamefont {Z.}~\bibnamefont {Song}}, \bibinfo {author} {\bibfnamefont
  {G.}~\bibnamefont {Chang}}, \bibinfo {author} {\bibfnamefont
  {I.}~\bibnamefont {Belopolski}}, \bibinfo {author} {\bibfnamefont
  {D.}~\bibnamefont {Multer}}, \bibinfo {author} {\bibfnamefont
  {M.}~\bibnamefont {Litskevich}}, \bibinfo {author} {\bibfnamefont {Z.-J.}\
  \bibnamefont {Cheng}}, \bibinfo {author} {\bibfnamefont {X.~P.}\ \bibnamefont
  {Yang}}, \bibinfo {author} {\bibfnamefont {B.}~\bibnamefont {Swidler}},
  \bibinfo {author} {\bibfnamefont {H.}~\bibnamefont {Zhou}}, \bibinfo {author}
  {\bibfnamefont {H.}~\bibnamefont {Lin}}, \bibinfo {author} {\bibfnamefont
  {T.}~\bibnamefont {Neupert}}, \bibinfo {author} {\bibfnamefont
  {Z.}~\bibnamefont {Wang}}, \bibinfo {author} {\bibfnamefont {N.}~\bibnamefont
  {Yao}}, \bibinfo {author} {\bibfnamefont {T.-R.}\ \bibnamefont {Chang}},
  \bibinfo {author} {\bibfnamefont {S.}~\bibnamefont {Jia}},\ and\ \bibinfo
  {author} {\bibfnamefont {M.}~\bibnamefont {Zahid~Hasan}},\ }\bibfield
  {title} {\bibinfo {title} {{Quantum-limit Chern topological magnetism in
  TbMn$_{6}$Sn$_{6}$}},\ }\href {https://doi.org/10.1038/s41586-020-2482-7}
  {\bibfield  {journal} {\bibinfo  {journal} {Nature}\ }\textbf {\bibinfo
  {volume} {583}},\ \bibinfo {pages} {533} (\bibinfo {year}
  {2020})}\BibitemShut {NoStop}%
\end{thebibliography}%
\end{document}


\widetext
\begin{center}
\textbf{\large Supplemental Material for ''Effect of magnetism and phonons on localized carriers in ferrimagnetic kagome metals \Gd\ and \Tb''}

\vspace{5mm}
M. Wenzel, A. A. Tsirlin, O. Iakutkina, Q. Yin, H.C. Lei, M. Dressel, and E. Uykur
\end{center}
\setcounter{equation}{0}
\setcounter{figure}{0}
\setcounter{table}{0}
\setcounter{page}{1}
\makeatletter
\renewcommand{\theequation}{S\arabic{equation}}
\renewcommand{\thefigure}{S\arabic{figure}}
\renewcommand{\bibnumfmt}[1]{[S#1]}
\renewcommand{\citenumfont}[1]{S#1}

\section{Crystal growth}
Single crystals of \Gd\ and \Tb\ were grown by the Sn flux method with Gd/Tb~:~Mn~:~Sn~=~1~:~6~:~20 molar ratio. Gd/Tb (ingots), Mn (pieces) and Sn (grains) were put into an alumina crucible and sealed in a quartz ampule under partial argon atmosphere. The sealed quartz ampule was heated up to 1373~K and kept there for 20~h to ensure the homogeneity of melt. After that, for \Gd, the temperature was rapidly cooled down to 1023~K for 20~h and subsequently cooling down to 873~K at 2~K/h. For \Tb, the temperature was cooled down directly to 873~K with the rate of 5~K/h. Finally, the ampules were taken out of furnace and the single crystals were separated from the flux by a centrifuge. 

\section{Experimental details}
Prior to our optical study, we carried out four-point dc resistivity and magnetic susceptibility measurements within the $ab$-plane to monitor possible magnetic transitions and confirm the stoichiometry. For the magnetic susceptibility measurements, a small magnetic field of $H$~=~0.1~T was applied. The obtained data agrees well with the literature and confirms the spin reorientation in \Tb\ around 310~K from the basal plane near to the $c$-axis \cite{Jones2022}. For \Gd, all magnetic transitions are above the measured temperature range; hence, we observed no anomalies in our data \cite{Gorbunov2012}.

\begin{figure}[b]
\centering
	\includegraphics[width=0.875\columnwidth]{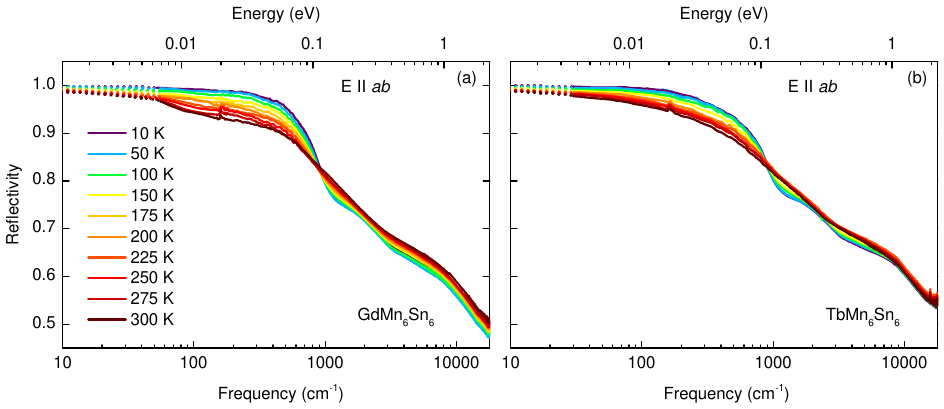}
	\caption{Temperature-dependent reflectivity over a broad frequency range (50 to 18000 \cm) measured in the $ab$-plane. The dotted lines are the Hagen-Rubens extrapolations.}		
	\label{reflectivity}
\end{figure}

Freshly cleaved samples with the dimensions of 2~x~2~mm$^2$ surface area and thickness of about 100~$\mu$m were used for the optical study. Here, temperature-dependent reflectivity measurements were performed in the $ab$-plane covering a broad frequency range from 50 to 18000 \cm\ (6.2~meV~-~2.23~eV) down to 10~K, as shown in Fig.~\ref{reflectivity}. For the high-energy range ($\omega$~$>$~600~\cm) a Bruker Vertex 80v spectrometer with an incorporated Hyperion IR microscope was used, while the low-energy range was measured with a Bruker IFS113v spectrometer and a custom-built cryostat. Freshly evaporated gold mirrors served as reference in these measurements. The absolute value of the reflectivity was obtained by an in-situ gold-overcoating technique in the far-infrared range, as described in ref.~\cite{Homes1993}

Considering the metallic nature of the samples, we used Hagen-Rubens extrapolation below 50~\cm, while x-ray scattering functions were utilized for the high-energy range to extrapolate the data \cite{Tanner2015}. The optical conductivity is then calculated from the measured reflectivity by standard Kramers-Kronig analysis.

\section{Decomposition of optical spectra}
Different contributions to the total optical conductivity were modeled with the Drude-Lorentz approach. With $\varepsilon_{\infty}$ being the high-energy contributions to the real part of the dielectric permittivity, the dielectric function [$\tilde{\varepsilon}=\varepsilon_1 + i\varepsilon_2$] is expressed as
\begin{equation}
\label{Eps}
\tilde{\varepsilon}(\omega)= \varepsilon_\infty - \frac{\omega^2_{p,{\rm Drude}}}{\omega^2 + i\omega/\tau_{\rm\, Drude}} + \sum\limits_j\frac{\Omega_j^2}{\omega_{0,j}^2 - \omega^2-i\omega\gamma_j}.
\end{equation}
Here, $\omega_{p,{\rm Drude}}$ and $1/\tau_{\rm\,Drude}$ are the plasma frequency and the scattering rate of the itinerant carriers, respectively. The parameters $\omega_{0,j}$, $\Omega_j$, and $\gamma_j$ describe the resonance frequency, width, and the strength of the $j^{th}$ excitation, respectively. 

\begin{figure}[b]
\centering
	\includegraphics[width=0.875\columnwidth]{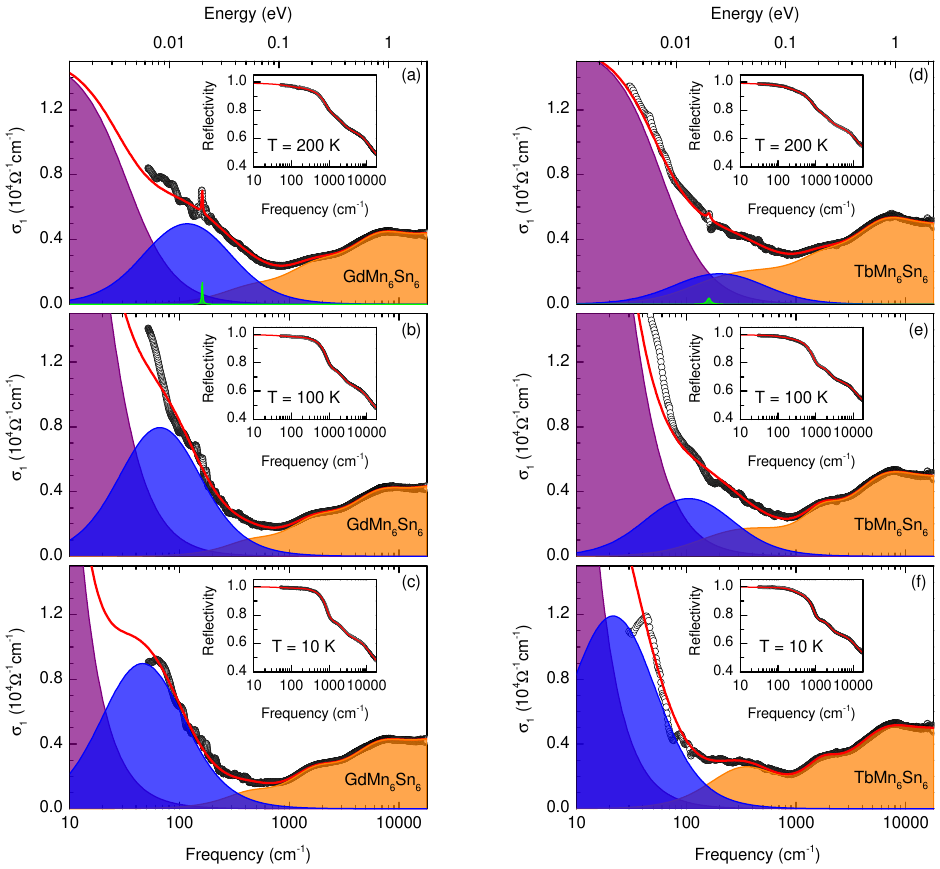}
	\caption{Decomposed optical conductivity at 200~K, 100~K, and 10~K, consisting of a Drude component (purple), a localization peak (blue), a phonon mode (green), and several interband transitions (orange) modeled with the Drude-Lorentz approach. The insets show the total fit to the measured in-plane reflectivity.} 	
	\label{decomp}
\end{figure}

\begin{table}
\begin{center}
{\begin{tabular} {c c c c c c c}
\toprule
\textit{Lorentzian 1}&\multicolumn{3}{c}{\Gd}& \multicolumn{3}{c}{\Tb}\\
 \midrule
$T$ (K)&$\Delta\varepsilon$  & $\omega_0$ (\cm)& $\gamma$ (\cm)&$\Delta\varepsilon$  & $\omega_0$ (\cm)& $\gamma$ (\cm)\\
10&172.705&432.556&651.473&909.009&332.377&682.769\\
50 &148.140	&432.556	&619.329&850.895&	332.377	&758.926 \\
100 &156.009&	432.556	&660.067& 760.626	&332.377	&899.605\\
150 &198.806	&432.556&	853.54&865.581&	332.377	&807.755\\
175 &189.058	&432.556	&901.554&-&-&-\\
200&202.635	&432.556	&767.635&1024.99&	332.377	&1041.35 \\
225 &215.642&	432.556	&1034.66&1050.23	&332.377	&982.126\\
250&192.204	&432.556&	1164.59&1180.61	&332.377&	1006.25 \\
275&163.49	&432.556	&1105.97&1265.02	&332.377	&1002.79\\
300&206.121	&432.556	&1215.73&1251.79	&332.377	&1049.3 \\
\bottomrule
\addlinespace[2pt]
\textit{Lorentzian 2}&\multicolumn{3}{c}{\Gd}& \multicolumn{3}{c}{\Tb}\\
 \midrule
$T$ (K)&$\Delta\varepsilon$  & $\omega_0$ (\cm)& $\gamma$ (\cm)&$\Delta\varepsilon$  & $\omega_0$ (\cm)& $\gamma$ (\cm)\\
10&95.7128	&1698.15	&2554.68&69.8246&	1705.89	&1924.48\\
50&99.9937	&1698.15	&2788.11&72.9802	&1705.89	&1985.85\\
100&100.325	&1698.15	&2776.85&76.0712&	1705.89	&2027.15\\
150&107.838	&1698.15	&2994.56&88.1694	&1705.89	&2441.02\\
175&110.115&	1698.15&	3048.64&-&-&-\\
200&113.452&	1698.15	&3017.55&98.043&	1705.89&	2775.34\\
225&108.198&	1698.15&	3016.64&103.049	&1705.89&	2831.12\\
250&126.412	&1698.15&	3437.89&93.1684	&1705.89	&2754.83\\
275&127.789	&1698.15	&3432.88&96.9267	&1705.89&	2883.27\\
300&121.526&	1698.15	&3432.53&100.372	&1705.89	&3007.04\\

\bottomrule
\addlinespace[2pt]
\textit{Lorentzian 3}&\multicolumn{3}{c}{\Gd}& \multicolumn{3}{c}{\Tb}\\
 \midrule
$T$ (K)&$\Delta\varepsilon$  & $\omega_0$ (\cm)& $\gamma$ (\cm)&$\Delta\varepsilon$  & $\omega_0$ (\cm)& $\gamma$ (\cm)\\
10&50.7642	&6479.28	&11939&74.6754	&6338.93&	12808.7\\
50&47.1342&	6479.28&	11939&75.5724	&6338.93	&12808.7\\
100&49.0074	&6479.28&	11939&76.3284&	6338.93&	12808.7\\
150&50.8744	&6479.28	&11939&76.336	&6338.93	&12808.7\\
175&50.9762	&6479.28	&11939&-&-&-\\
200&53.1189	&6479.28	&11939&77.1148	&6338.93	&12808.7\\
225&53.6501	&6479.28	&11939&78.7435	&6338.93	&12808.7\\
250&52.5772	&6479.28&	11939&76.814	&6338.93	&12808.7\\
275&53.1029	&6479.28	&11939&76.8908&	6338.93&	12808.7\\
300&55.8173	&6479.28&	11939&73.3722	&6338.93	&12808.7\\
\bottomrule
\addlinespace[2pt]
\textit{Lorentzian 4}&\multicolumn{3}{c}{\Gd}& \multicolumn{3}{c}{\Tb}\\
 \midrule
$T$ (K)&$\Delta\varepsilon$  & $\omega_0$ (\cm)& $\gamma$ (\cm)&$\Delta\varepsilon$  & $\omega_0$ (\cm)& $\gamma$ (\cm)\\
10&18.9776	&31246.3	&80843.9&19.2266	&27418.1&	61229.7\\
50&19.2876	&31246.3	&80843.9&19.0763	&27418.1	&61229.7\\
100&19.3669	&31246.3	&80843.9&18.9045	&27418.1	&61229.7\\
150&19.1541&	31246.3	&80843.9&19.0954	&27418.1	&61229.7\\
175&19.0385	&31246.3	&80843.9&-&-&\\
200&19.0366&	31246.3&	80843.9&18.7211	&27418.1	&61229.7\\
225&18.8462	&31246.3	&80843.9&18.9102	&27418.1&	61229.7\\
250&19.0347	&31246.3	&80843.9&18.7916	&27418.1	&61229.7\\
275&19.029	&31246.3	&80843.9&18.7916	&27418.1&	61229.7\\
300&19.2212&	31246.3	&80843.9&19.0186	&27418.1&	61229.7\\

\bottomrule
\end{tabular}}
\caption{Fit parameters of the total number of four Lorentzians used to model the interband optical transitions in \Gd\ and \Tb. \label{Lorentzians}}
\end{center}
\end{table}

Following the approach of previous optical studies of kagome metals, we base our analysis of the localization peak on the displaced Drude formalism proposed in 2014 by Fratini \textit{et al.} \cite{Fratini2014}. Here, possible localization effects, due to interactions of charge carriers with low-energy degrees of freedom, such as phonons, electric or magnetic fluctuations, are considered by modifying the classical Drude response with an additional backscattering of the electrons. This leads to a shift of the zero-frequency response to a finite value:
\begin{equation}
\tilde{\sigma}_{\rm localization}(\omega)=\frac{C}{\tau_{\mathrm{b}}-\tau}\frac{\tanh\{\frac{\hbar\omega}{2k_{\mathrm{B}}T}\}}{\hbar\omega}\,\cdot\,\mathrm{Re}\left\{\frac{1}{1-\mathrm{i}\omega\tau}-\frac{1}{1-\mathrm{i}\omega\tau_{\mathrm{b}}}\right\}.
\end{equation}
Here, $C$ is a constant, $\hbar$ is the reduced Planck constant, $k_{\mathrm{B}}$ the Boltzmann constant, $\tau_{\mathrm{b}}$ the backscattering time, and $\tau$ the elastic scattering time of the standard Drude model. 

The total dielectric permittivity takes the form
\begin{equation}
\tilde{\varepsilon}(\omega) = \tilde{\varepsilon}_{\rm Drude}(\omega) + \tilde{\varepsilon}_{\rm Lorentz}(\omega)+\tilde{\varepsilon}_{\rm localization}(\omega).
\end{equation}
The complex optical conductivity [$\tilde{\sigma}=\sigma_1 + i\sigma_2$] is then calculated as
\begin{equation}
\tilde{\sigma}(\omega)= -i\omega[\tilde{\varepsilon} (\omega) -\varepsilon_\infty]/4\pi.
\label{Cond}
\end{equation}

Fig.~\ref{decomp} shows the decomposed optical conductivity at various temperatures. The spectra were fitted in a consistent way for all temperatures using one Drude contribution (purple), a total number of four Lorentzians (see Table~\ref{Lorentzians} for the parameters) to describe the interband optical transitions (orange), a sharp Lorentzian for the phonon mode (green), as well as the Fratini model to describe the localization peak (blue). At 300~K the localization peak is only weakly pronounced and additionally screened by low-energy interband transitions in \Tb. On the other hand, the peak is clearly visible by the eye in the spectrum of \Gd\ due to the absence of strong low-energy interband absorptions and the sharper Drude contribution.

\begin{figure}
\centering
	\includegraphics[width=1\columnwidth]{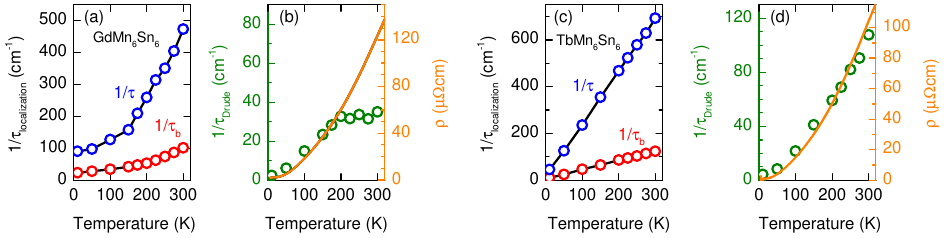}
	\caption{Elastic scattering rate, 1/$\tau$ (blue) and backscattering rate, 1/$\tau_{\mathrm{b}}$ (red) of the Fratini model fits. Additionally, the elastic scattering rate
of the Drude contribution (green), overlaid with the dc resistivity (orange), is given.}		
	\label{scattering}
\end{figure}

In Fig.~\ref{scattering}, we show the elastic scattering rate and the backscattering rate obtained from the Fratini model fits to the optical spectra, as well as the scattering rate of the classical Drude model. When overlaying the Drude scattering rate with the dc resistivity, a remarkably similar temperature evolution is found in \Tb, indicating that the dc transport is governed by the free electrons. On the other hand, a clear deviation of this behavior above $\sim$~200~K is observed in \Gd. Considering the akin temperature dependence of the resistivity to the elastic scattering rate of the localization peak at high temperatures, this signals a significant contribution of the incoherent carriers to the dc transport in \Gd. 

\section{Phonon calculations}
Phonon calculations were performed on the density-functional theory (DFT) level in \texttt{VASP}~\cite{vasp1,vasp2} using the refined structural parameters given in Table~\ref{XRD} and the Perdew-Burke-Ernzerhof (PBE) flavor of the exchange-correlation potential~\cite{pbe96}. Spin-orbit coupling was included, and different directions of the magnetic moment were chosen. Ferromagnetic order was introduced for Mn atoms, whereas $f$-electrons of Gd and Tb were placed into the core, and only a small residual magnetic moment due to $d$-electrons appeared on these atoms. This simplification was necessary in order to achieve good convergence of total energies and forces, as required in phonon calculations. The $8\times 8\times 4$ $k$-mesh was used.

\begin{figure}[h]
        \centering
       \includegraphics[width=1\columnwidth]{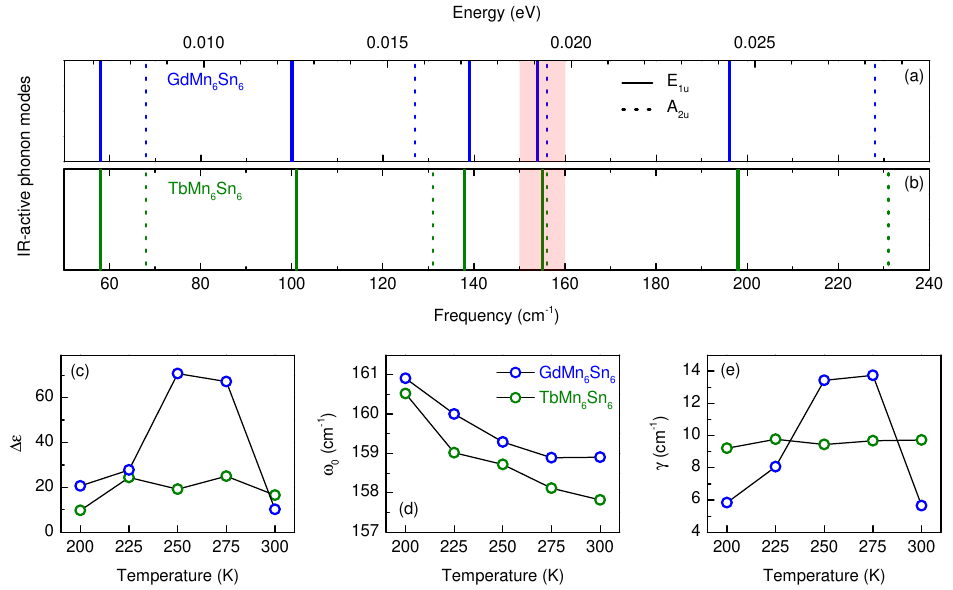}
    
   \caption{(a) and (b) Calculated IR-active phonon frequencies of \Gd\ and \Tb. The solid lines represent the in-plane E$_{1u}$ modes while dashed lines mark A$_{2u}$ modes involving atomic displacements along the $c$-axis.(c)-(e) Fit parameters of the observed phonon mode in the optical spectra corresponding to the  E$_{1u}$ mode marked by the red area in (a) and (b).}	
    \label{phonon}
\end{figure}

Frequencies of $\Gamma$-point phonons were obtained from the built-in procedure with frozen atomic displacements of 0.015~\AA. Fig.~\ref{phonon} (a) and (b) depict the calculated IR-active phonon modes of \Gd\ and \Tb. In both compounds, a total number of nine IR-active modes are expected, which do not significantly vary in frequency with changes in the direction of the magnetic moments. Four of these are A$_{2u}$ $c$-axis modes (dashed lines) and hence, cannot be observed in our in-plane measurements. The remaining five modes are E$_{1u}$ modes (solid lines) involving in-plane atomic displacements. However, the appearance of phonon modes in reflectivity spectra strongly depends on the intensity of the phonon mode, especially for highly metallic samples, as in the case of the \Re\ series. Hence, it is possible that only the E$_{1u}$ mode around 160~\cm\ is strong enough to be captured by our measurements. This mode can be represented with a sharp Lorentzian,
\begin{equation}
\sigma_{1}(\omega) = \frac{\Delta \varepsilon \omega^2\omega_{0}^2\gamma}{4\pi[(\omega^2 - \omega_{0}^2)^2 + \gamma^2\omega^2]} .
\end{equation}
Here, $\Delta\varepsilon$ stands for the intensity, $\omega_0$ for the resonance frequency, and $\gamma$ for the linewidth. Consistent with the hardening of the lattice, we observe a slight blue shift of the mode upon cooling in both compounds. In \Gd, a significant enhancement of intensity and a slight broadening of the mode are observed as the localization peak crosses the respective phonon mode. On the other hand, no such changes are observed in \Tb. Here, both the intensity as well as the linewidth stay constant within the error bars of our fits.

\begin{table}
   
    \begin{minipage}{0.49\linewidth}
        \centering
        
        {\begin{tabular} {c c c c c}
\multicolumn{5}{c}{\textbf{\Gd}}\\\hline
\multicolumn{5}{c}{$a=b=5.5399(2)$ \AA, $c=9.0318(5)$ \AA}\\
\multicolumn{5}{c}{$V=240.054(18)$ \AA$^3$}\\
\multicolumn{5}{c}{P6/$mmm$}\\
\multicolumn{5}{c}{$\lambda=0.71073$ \AA}\\
\multicolumn{5}{c}{$\Theta_{\mathrm{min}}=0.41^{\circ}$, $\Theta_{\mathrm{max}}=27.48^{\circ}$}\\
\multicolumn{5}{c}{$-7\leq h \leq 7$, $-7\leq k \leq 7$, $-11\leq l \leq 11$}\\
\multicolumn{5}{c}{$R_{\mathrm{int}}=0.0656$}\\
\midrule
Atom&$x/a$&$y/b$&$z/c$&$U_{\mathrm{iso}}$ (\AA$^2$)\\
Gd&0&0&0.5&0.01120(34)\\
Mn&0.5&0&0.25224(13)&0.01074(34)\\
Sn1&$\frac{1}{3}$&$\frac{2}{3}$&0&0.01147(33)\\
Sn2& $\frac{1}{3}$&$\frac{2}{3}$&0.5&0.01024(33)\\
Sn3&0&0&0.16206(11)&0.01184(33) \\
\bottomrule
\end{tabular}}

    \end{minipage}
    \begin{minipage}{0.49\linewidth}
         \centering
        
        {\begin{tabular} {c c c c c}
\multicolumn{5}{c}{\textbf{\Tb}}\\\hline
\multicolumn{5}{c}{$a=b=5.5305(2)$ \AA, $c=9.0223(5)$ \AA}\\
\multicolumn{5}{c}{$V=238.988(18)$ \AA$^3$}\\
\multicolumn{5}{c}{P6/$mmm$}\\
\multicolumn{5}{c}{$\lambda=0.71073$ \AA}\\
\multicolumn{5}{c}{$\Theta_{\mathrm{min}}=0.41^{\circ}$, $\Theta_{\mathrm{max}}=27.48^{\circ}$}\\
\multicolumn{5}{c}{$-6\leq h \leq 7$, $-7\leq k \leq 7$, $-11\leq l \leq 11$}\\
\multicolumn{5}{c}{$R_{\mathrm{int}}=0.0787$}\\
\midrule
Atom&$x/a$&$y/b$&$z/c$&$U_{\mathrm{iso}}$ (\AA$^2$)\\
Tb&0&0&0.5&0.01182(41)\\
Mn&0.5&0&0.25244(16)&0.01129(42)\\
Sn1&$\frac{1}{3}$&$\frac{2}{3}$&0&0.01155(40)\\
Sn2& $\frac{1}{3}$&$\frac{2}{3}$&0.5&0.01078(40)\\
Sn3&0&0&0.16276(13)&0.01211(40) \\
\bottomrule
\end{tabular}}
    \end{minipage}
\caption{Details of data collection and refined structural parameters for \Gd\ (left) and \Tb\ (right).}
    \label{XRD}
\end{table}

\section{Calculation of the optical conductivity}
DFT calculations of the band structure and optical conductivity were performed in the \texttt{Wien2K} code~\cite{wien2k} using the same PBE functional~\cite{pbe96}. Spin-orbit coupling was included in all calculations. For a realistic implementation of the magnetic structures, the [100]-direction of the magnetic moments was chosen for \Gd, while the [001]-direction was set for \Tb. Additionally, an antiferromagnetic coupling between the Mn- and rare-earth-atoms was implemented. Moreover, a Hubbard $U_{Gd/Tb}$~=~10~eV was added to the $4f$ shell of the rare-earth element using the DFT+$U$ method with the FLL (fully localized limit) double-counting correction to push the minority 4$f$ states to energies well above the Fermi level. DFT calculations were converged on the $15\times 15\times 4$ $k$-mesh. Optical conductivity was calculated within \texttt{Wien2K}~\cite{draxl2006} on a denser $26\times 26\times 14$ $k$-mesh. 

\begin{figure}[h]
\centering
	\includegraphics[width=0.875\columnwidth]{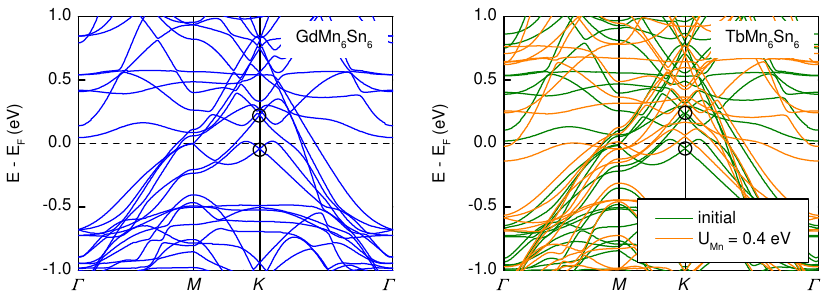}
	\caption{DFT+$U$ band structures for \Gd\ (left) and \Tb\ (right) shown along high-symmetry paths of the first Brillouin zone. The observed Dirac points at the $K$ point are marked with circles and their energies are noted in Table~\ref{Diractable}.}		
	\label{bands}
\end{figure}

Fig.~\ref{bands} shows the calculated band structures along high-symmetry paths of the first Brillouin zone. Both compounds possess flat bands around 0.5 eV and saddle points nearby the $M$ point. The Dirac points nearby $K$ are marked by circles, and their energies are noted in Table~\ref{Diractable}. In the case of a two-dimensional Dirac point, the optical conductivity is supposed to show a sharp Drude component along with a step-like onset at 2$|E_{\mathrm{D}}|$, followed by a frequency-independent behavior. Hence, the interpretation of the observed steps in the optical conductivity as the signature of two-dimensional Dirac fermions is very tempting. The obtained Dirac cone energies from our experiment are noted in Table~\ref{Diractable}. A direct comparison with our calculations reveals a remarkable agreement of the determined energies. On the other hand, a comparison with ARPES studies shows a larger deviation of the energies of the second Dirac point. However, it should be noted that
ARPES only probes the states below the Fermi energy leading to less accurately determined values. 

\begin{figure}
\centering
	\includegraphics[width=0.875\columnwidth]{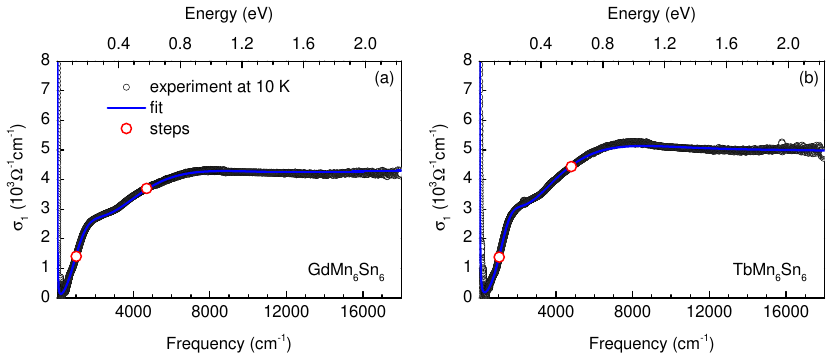}
	\caption{Experimental optical conductivity after subtracting the localization peak and the low-energy interband transitions. The remaining spectra resemble the optical conductivity of two-dimensional Dirac fermions. The steps at 2$|E_{\mathrm{D}}|$ are highlighted with dots.}		
	\label{dirac}
\end{figure}

Despite the good agreement between the experiment and our calculations, the step-like absorption features shown in Fig.~\ref{dirac} should be interpreted cautiously. A closer look at the calculated bandstructure reveals the large number of bands crossing the Fermi energy in these compounds. Thus, the multi-band nature of the \Re\ series should not be disregarded.

For \Tb, the accuracy of the calculated optical conductivity in comparison with the experiment increases when either shifting the Fermi energy down by 47~meV or adding a Hubbard $U_{\mathrm{Mn}}$~=~0.4~eV to the Mn-atoms. However, in both cases, the energy of the first Dirac point shifts above the Fermi level, which is not expected from ARPES studies on the \Re\ series.

\begin{table}[h]
\begin{center}
\begin{tabular} {c|c|c|c|c|c|c}
&\multicolumn{2}{c|}{optical study}  & \multicolumn{2}{c|}{calculations} & \multicolumn{2}{c}{ARPES estimates} \\
 & $E_{\mathrm{D1}}$ (meV)  &$E_{\mathrm{D2}}$ (meV) &$E_{\mathrm{D1}}$ (meV)&$E_{\mathrm{D2}}$ (meV)& $E_{\mathrm{D1}}$ (meV)  &$E_{\mathrm{D2}}$ (meV) \\
    \hline
\Gd\ &63&291&- 42&233&- 42 \cite{Liu2021} &170 \cite{Ma2021a}\\
\Tb\ &65&298&- 41&239&not reported&130 \cite{Yin2020}
\end{tabular}
\caption{Energies of the Dirac points obtained from the optical study at $T\,=\,10\,\mathrm{K}$, the DFT+$U$ calculations, and estimates from ARPES measurements. \label{Diractable}}
\end{center}
\end{table}

\bibliography{TbGdsup}